\documentclass[final,twocolumn]{elsarticle}
\usepackage[T1]{fontenc}
\usepackage[latin9]{inputenc}
\usepackage{geometry}
\usepackage{textcomp}
\usepackage{amsmath}
\usepackage{amssymb}
\usepackage{graphicx}
\usepackage{gensymb}

\makeatletter
\def\ps@pprintTitle{%
 \let\@oddhead\@empty
 \let\@evenhead\@empty
 \def\@oddfoot{}%
 \let\@evenfoot\@oddfoot}

\providecommand{\tabularnewline}{\\}

\usepackage{babel}

\makeatother

\begin{document}
\begin{frontmatter}
\title{A novel approach to the localization and the estimate of radioactivity
in contaminated waste packages via imaging techniques}


\author[myaffiliation1,myaddress1]{Michele Pastena\corref{mycorrespondingauthor}}
\cortext[mycorrespondingauthor]{Corresponding author Tel. +49 (0) 6221 / 65175138}
\ead{michele.pastena@safetec-hd.de}
\author[myaffiliation1,myaddress1]{Bastian Weinhorst}
\author[myaddress2]{G{\"u}nter Kanisch}
\author[myaffiliation1,myaddress1]{Edgar P{\'e}rez Lezama}
\author[myaffiliation1,myaddress1]{Johannes Radtke}
\author[myaffiliation1,myaddress1]{Tim Thomas\corref{mycorrespondingauthor3}}

\address[myaffiliation1]{Safetec Entsorgungs- und Sicherheitstechnik GmbH}
\address[myaddress1]{Kurpfalzring 98a, D-69123 Heidelberg, Germany}

\address[myaddress2]{Wittland 44g, D-22589 Hamburg, Germany}

\begin{abstract}
Dismantling nuclear power plants entails the production of a large amount of contaminated (or potentially contaminated) waste that must be disposed according to national and international regulations. A large part of the end products needs to be stored in special repositories, but a significant part of it is slightly contaminated or not contaminated at all, making it possible to free release it. One possible approach to free release measurements uses Large Clearance Monitors, chambers surrounded by plastic scintillation detectors that can measure up to 1000kg of waste. Due to the composite nature of the detection system in a Large Clearance Monitor, it is easy to imagine that one can apply 3D imaging algorithms to localize radioactive sources inside a waste package. In this work we will show how a special algorithm that maximizes the conditional informational entropy allows decisions about the clearance of portions of the sample.
\end{abstract}


\end{frontmatter}

\section*{Introduction}

One major path for clearance of large amount of waste arising from
the decommission of nuclear power plants is based on measurements with
Large Clearance Monitors (LCM), devices that utilize multi-detector
plastic scintillation arrays (LCMs can
consist of up to 24 plastic scintillation detectors) to offer both,
extremely low minimum detectable activities (MDA) and short measurement
times for waste packages weighing up to one metric ton. LCMs represent in
general the most efficient way to measure big and heavy
samples. Such clearance measurements are performed also by means germanium detectors  (or arrays of germanium detectors) with beam collimators both for waste packages
(rotating the sample and measuring its emitted radiation in several
different directions) and for whole rooms or buildings (measuring the
radiation emitted by the surface of a room by means of a collimated
static detector). It has to be pointed out though that germanium detectors
are slow and inefficient as each single measurement requires
25-30 minutes and furthermore such devices require delicate cooling systems 
with liquid nitrogen. In order to handle the large amount of waste
produced during the decommissioning of a nuclear power plant, the measurement for the release procedure must be sufficiently short. For this reason, Large Clearance Monitors (LCM) are used for this task.

However, a major problem with these devices is to obtain reliable
information of so-called hotspots, i.e. contaminated or activated
material in the waste-package. While the $4\pi$-setup of the detectors
allows a crude localization of one hotspot (Mirion Technologies RTM644Inc
User Manual), no analysis has been conducted so far to test the uncertainty
of the estimated activity. Moreover, the localization of hotspots are routinely
implemented in a sufficiently high radioactive environment \cite{intro:14,intro2:15}
by employing high-purity germanium detectors, but the low radioactivity
needed for clearance measurements, presents a different problem. Rather
than posing the question where the activity is likely to be located
for signals which can be clearly distinguished from the background, for
clearance measurements all possible geometrical and hardware uncertainties
have to be taken into account for the minimum detection limit and
thus the possible activity distributions have to be derived according
to ISO 11929. In order to solve the problem of distinguishing
the background radiation from the one produced by a sample with a
contaminated hotspot, we propose a new method that resembles the imaging
techniques used in medical physics.

In this paper we want to present a novel approach to the localization
and the estimate of radioactivity in contaminated waste packages.
The new procedure uses an imaging algorithm to reconstruct accurate
images of the radioactivity distribution in a given waste package,
allowing fast and reliable decisions for clearance of potentially
contaminated or activated material. The proposed method relies on
the maximization of informational entropy. This algorithm is used
in the medical field for the imaging of tissues in the human body
by means of PET/SPECT scans. 

In the section \ref{sec:clearance_now} we will summarize the most common
techniques used to determine the activity of low activated material
that can be potentially free released (clearance measurements) mentioning
the models and techniques prescribed by the international standards.
In the sections from \ref{sec:CEM} to \ref{sec:cem_fma} we will
present the Conditional Entropy Maximization, an imaging technique
widely used in medical physics and we will show the changes needed
to adapt it to the purpose of clearance measurements. In section \ref{sec:characteristic_limits}
we will present the mathematical machinery needed to make the aforementioned algorithm compliant with the
regulatory norms regarding the determination of the characteristic
limits (decision threshold, detection limit and limits of the coverage
interval) for measurements of ionizing radiation provided by the international
standard. In the final section \ref{sec:examples} we will apply
the Conditional Entropy Maximization to some simulated samples.
This research did not receive any specific grant from funding agencies in the public, commercial, or
not-for-profit sectors.

\section{\label{sec:clearance_now}State of art of the clearance measurements}

Usually a LCM consists of a chamber encompassed by a gamma ray detection
system distributed over the whole solid angle around the sample, consisting
of 24 polyvinyl toluene (PVT)-based large volume plastic scintillation
detectors. In order to reduce the background radiation, massive shielding
(i.e. 130 mm steel) surrounds the chamber. The activity of the waste
is then estimated using a combined measurement of the 24 count rates.
This way the LCM can measure up to 1000 kg of waste in one minute.
The LCM is also equipped with a conveyor system for the waste packages.

\begin{figure}
\begin{centering}
\includegraphics{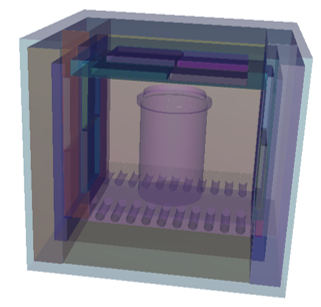} 
\par\end{centering}
\caption{Reproduction of the chamber of a Large Clearance Monitor}
\end{figure}

One major drawback of the plastic scintillation detectors used in
the LCMs is the poor energy resolution that makes the identification
of the radionuclides in the waste almost impossible. This means that gamma
spectra cannot be identified, but only the total count in a given
time (usually 60 s) is measured for a really wide range of energies.
The estimate of the activity depends then on the knowledge of the
radionuclide inventory, also known as nuclide vector (NV). Furthermore,
the lacking identification of gamma lines does not allow to discriminate
between natural radio nuclides and those produced during the operation
of the nuclear power plant.

To obtain the activity of a waste package, the total count rate and
a calibration factor $w$ are needed. The factor $w=1/\varepsilon$
is the inverse of the detection efficiency of the whole system (waste
and measurement device) i.e. the ability of detecting the photons
emitted by the radioactive sources, and depends on the composition
of the nuclide inventory. However, the absolute efficiency depends
not only on the detection properties of the detector (also known as
detector characterization), but is also determined by the waste package,
material, and activity distribution (referred to as geometry).

\subsection{\label{subsec:DIN_ISO}ISO 11929:2019}

The aforementioned calibration factor $w$ is one of the quantities
needed to estimate the activity of a waste package. A physical model
of the connection between the activity and the other physical observables
(count of gamma photons, shielding) is suggested by the international
standard ISO 11929 for each type of measurement and environment (high
activity, low activity, clearance measurement, etc.). The reason of
the importance of this international standard lays on the fact that
the law in the field of nuclear safety prescribes the adoption of
state-of-art models, techniques and technologies that make the handling
of radioactive sources as safe as possible. The duty of the ISO 11929
standard is then to summarize all the most recent and efficient models
and prescriptions to be adopted. With reference to the ISO 11929:2019
standard, from now on we will take care exclusively of clearance measurements
(i.e. measurements of samples with such a low activity that they are potentially
free releasable). Once we know the calibration factor $w$ of the
measurement device that we are using (both plastic and germanium detectors),
the simplified model to evaluate the activity $a$ is 
\begin{equation}
a=\left(r_{g}-r_{0}\cdot x_{3}-x_{41}+x_{42}\right)w\label{eq:eval_model}
\end{equation}

with 
\begin{itemize}
\item $r_{g}$ gross count rate measured with the sample in the chamber, 
\item $r_{0}$ background count rate measured with empty chamber, 
\item $x_{3}$ correction of the background count rate for its variability
due to work activities near the device for clearance measurements, 
\item $x_{41}$ correction for natural radionuclides, i.e. $^{40}K$ and
the $^{226}Ra$ and $^{232}Th$ decay series, in the material to be
measured, 
\item $x_{42}$ correction for shielding of the background by the material
to be measured. 
\end{itemize}
It has to be remarked that in the present work we will be analyzing
only $^{60}Co$ sources that emit two photons per decay with a probability close to 1. In general it can happen that the nuclide inventory (i.e. the type and relative percentage of radionuclides) is much more complex, requiring more attention in the formulation and solution of the problem. In order to lower the external dose of ionizing radiation, international norms prescribe to fully characterize
the activity distribution of a sample (e.g. a waste package) to be
analyzed. The core idea is that it is not sufficient that the measurand
is below the prescribed threshold, but that we also need a high statistical
confidence (typically 95\%). This can be achieved either by comparing
case by case the whole statistical distribution with the prescribed
thresholds or by calculating the characteristic values (in particular
the upper limit of the coverage interval and the detection limit).
The second option is clearly the most straightforward, and to this
end we need the uncertainty and all the statistical characteristic
values according to \cite{GUM:9}. The uncertainty for the activity
(\ref{eq:eval_model}) is given by 
\begin{align*}
u^{2}\left(a\right) & =w^{2}\left[\frac{r_{g}}{t_{g}}+x_{3}^{2}\frac{r_{0}}{t_{0}}+r_{0}^{2}x_{3}^{2}u^{2}\left(x_{3}\right)\right.\\
 & \left.+u^{2}\left(x_{41}\right)+u^{2}\left(x_{42}\right)\right]+a^{2}u_{rel}^{2}\left(w\right)
\end{align*}

with $u_{rel}\left(w\right)=u\left(w\right)/w$ relative uncertainty
of $w$. This formula is the starting point to calculate all the characteristic
limits of the activity distribution, as it is explained in
\ref{sec:appendix_characteristic_limits}. Indeed, the uncertainty
for an assumed true value $\tilde{a}$ is then 
\begin{multline*}
\tilde{u}^{2}\left(\tilde{a}\right)=w^{2}\left[\frac{1}{t_{g}}\left(\frac{\tilde{a}}{w}+\frac{n_{0}}{t_{0}}x_{3}+x_{41}+x_{42}\right)+x_{3}^{2}\frac{r_{0}}{t_{0}}\right.\\
\left.+r_{0}^{2}\cdot u^{2}\left(x_{3}\right)+u^{2}\left(x_{41}\right)+u^{2}\left(x_{42}\right)\right]\\
+\tilde{a}^{2}u_{rel}^{2}\left(w\right)
\end{multline*}

and thus we can calculate the following characteristic values for
the simplified evaluation model: 
\begin{itemize}
\item decision threshold
\begin{multline*}
a^{*}=k_{1-a}\tilde{u}\left(0\right)=\\
k_{1-\alpha}w\left\{ \frac{1}{t_{g}}\left(\frac{n_{0}}{t_{0}}x_{3}+x_{41}+x_{42}\right)+x_{3}^{2}\frac{r_{0}}{t_{0}}\right.\\
\left.+r_{0}^{2}u^{2}\left(x_{3}\right)+u^{2}\left(x_{41}\right)+u^{2}\left(x_{42}\right)\right\} ^{1/2}
\end{multline*}
where $k_{1-\alpha}$ is the $\left(1-\alpha\right)$-quantile of
a normal distribution (typically $\alpha=0.05$), 
\item detection limit: it is obtained by solving the recursive equation
\begin{equation}
a^{\#}=a^{*}+k_{1-\beta}\tilde{u}\left(a^{\#}\right)\label{eq:detection_limit}
\end{equation}
where $k_{1-\beta}$ is the $\left(1-\beta\right)$-quantile of a
normal distribution (typically $\beta=0.95$), 
\item upper and lower limit of the coverage interval 
\begin{eqnarray*}
a^{\triangleleft}=a-k_{p}u\left(a\right) & ; & a^{\triangleright}=a+k_{q}u\left(a\right)
\end{eqnarray*}
where $p=\omega\left(1-\gamma/2\right)$, $q=1-\omega\gamma/2$, with
$\omega=\Phi\left(y/u\left(y\right)\right)$ probability corresponding
to $y/u\left(y\right)$ (assuming that the activity is normally distributed),
and $\gamma=0.05$. 
\end{itemize}
These definitions hold in the case of normally distributed activities.
In the most general case, we do not need to specify a probability
distribution as it is discussed in \ref{sec:appendix_characteristic_limits}.
As it has been pointed out, these values allow us to make decisions
about the riskiness of contaminated and nuclear waste in the case
of clearance measurements.

\section{\label{sec:CEM}Conditional Entropy Maximization for contaminated
waste}

In the field of medical physics, reconstructing images of the inner
parts of a human body for clinical purposes has always been a challenge.
The Single Photon Emission Tomography (SPECT) is mostly used to detect
functional or metabolic images of a tissue \cite{max_likelihood}
and has analogies with the characterization of radioactive waste packages.
The core idea of the SPECT in medical physics is to inject a radioactive
substance (typically emitting X rays) in living tissues and measure
the count rate of photons out of the body along several different directions.
This is achieved by encompassing the body in a 360\textdegree\, (in a plane) or
a $4\pi$ (in the space) detection system: this allows the reconstruction
of the tissue itself. A similar imaging technique is the Positron
Emission Tomography (PET) in which, instead, a positron-electron pair
annihilates and emits two photons in opposite directions \cite{kak-slaney:11};
in this case the image is reconstructed via measurements of the coincidence
count rate. A plethora of imaging algorithms has been proposed \cite{survey_med_im}
in the past decades; here we will focus on the Conditional Entropy
Maximization (CEM), in which the image of a radioactive or contaminated
source is reconstructed by means of the maximization of the informational
entropy.

Entropy maximization is a general technique to solve problems whose
analytic solution is not possible or computationally complicated.
This approach requires a straightforward and unambiguous definition
of the entropy \cite{entropy_def_1:3,entropy_def_2:4}.
Entropy maximization overcomes issues of other methods and furthermore
allows to bring prior information into the calculation.

The general idea behind the entropy maximization is to define a space
$Y$ of the measured observable $y\in Y$ (a single quantity or a
vector of quantities) and a parameter space $\Lambda$ of the observable
$\lambda\in\Lambda$ to be estimated. We can then define a probability
density function $P\left(y;\lambda\right)$ over the parameter space
that is in general difficult to maximize (i.e. we cannot find in general the most 
likely $\lambda$ that realizes the observed $y$). Therefore, we introduce
a fictitious space $X$ and a mapping $x\rightarrow y$, and thus we move
our problem of finding the parameter value (or values) to the problem
of maximizing another functional (in this case the conditional entropy)
defined on a different space. The task is then to assign an entropy
function that is compatible with the prior information about the system
and the measurement process, and then maximize it with a constraint
given by the outcome of the measurement.

\begin{figure}
\begin{centering}
\includegraphics[scale=0.4]{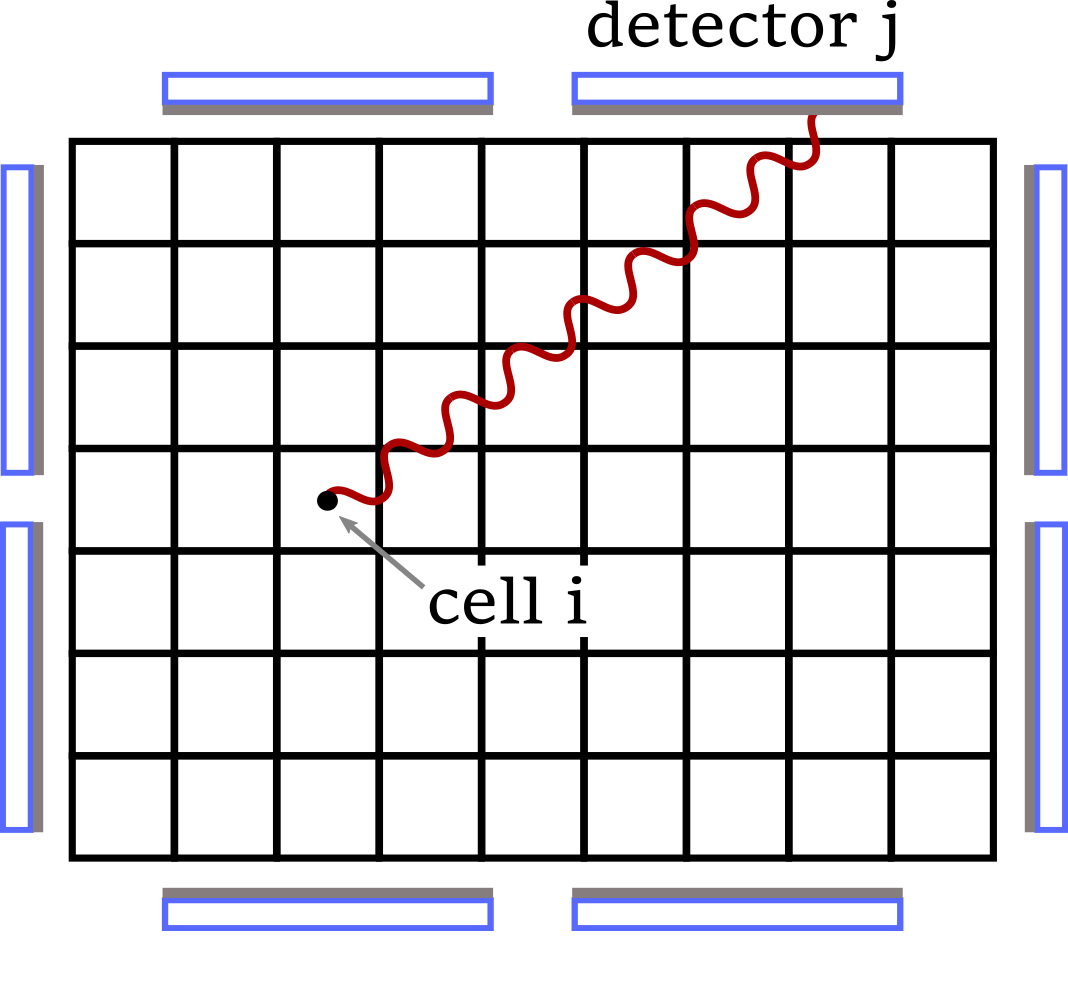} \includegraphics[scale=0.3]{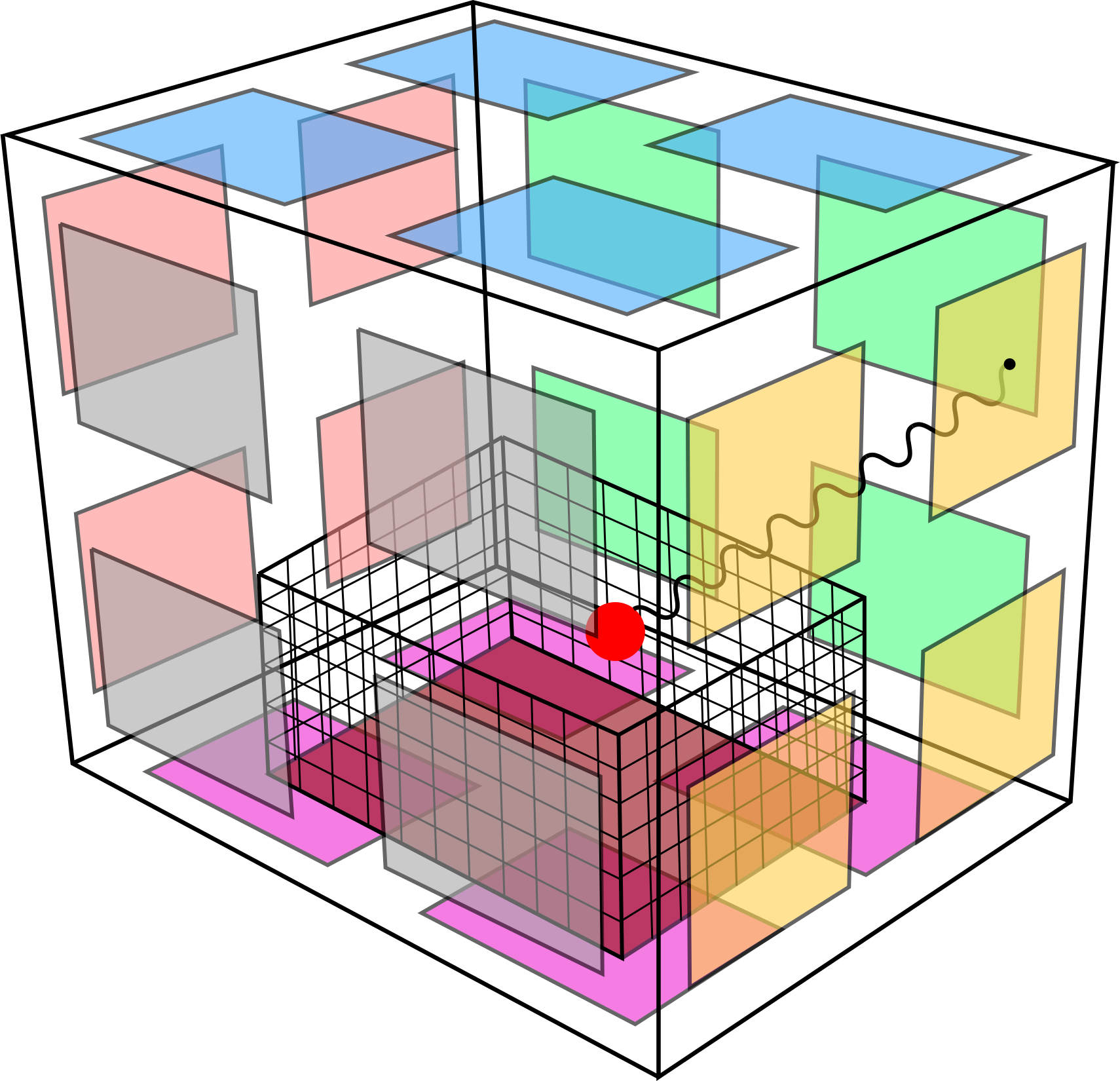} 
\par\end{centering}
\caption{\label{fig:model_SPECT}Model of the measurement process and of the
sample. In the top picture it is possible to see a section of the sample and of the LCM showing how
the emission and detection is modeled in order to properly define
the conditional entropy and the detection efficiency. In the bottom picture
an illustrative 3D model of the LCM and of the measurement process.}
\end{figure}

We will now show how this general approach can be applied to the
reconstruction of an image by measuring the count of emitted photons
(or coincidence counts). The measurement is performed by enclosing
the sample in a chamber surrounded by detectors (in a plane or in
the space) that can measure the number of emitted photons in different
directions (see Fig. \ref{fig:model_SPECT}). The measured count rates
$y_{j}$ ($j=1,\ldots,M$ with $M$ total number of detectors) can
be modeled as a Poisson variable and the
sample as a matrix of $N$ pixels (or cells), each
with activity $\text{\ensuremath{\lambda_{i}}}$. The goal is thus
to properly model the emission and measurement process in order to
define the probability (or likelihood) function $P\left(y;\lambda\right)$.
First of all, we can define a matrix $p_{ij}$ of the ratios between
the intensity detected by detector $j$ and the decay rate at
cell (i.e. pixel) $i$. In the following we will refer to these quantities as efficiencies.
The mean value of the measured count is then given by 
\begin{equation}
\mu_{j}=t_{m}\sum_{i=1}^{N}\lambda_{i}p_{ij}\label{eq:mean_value_poissonian}
\end{equation}
with $t_{m}$ time of the measurement. If we assume the measured counts to be Poisson distributed, then the conditional probability
that a realization \mbox{$y=\left(y_{1},\ldots,y_{M}\right)$} of the measurand
is obtained given a set of $\lambda_{i}$ is the joint distribution
of $M$ Poisson variables 
\[
P\left(y|\lambda\right)=\prod_{j=1}^{M}\frac{e^{-\mu_{j}}\mu_{j}^{y_{j}}}{y_{j}!}
\]
that takes the following form by inserting the (\ref{eq:mean_value_poissonian})
\[
P\left(y|\lambda\right)=\prod_{j=1}^{M}\frac{\exp(-t_{m}\sum_{i=1}^{N}\lambda_{i}p_{ij})(t_{m}\sum_{i=1}^{N}\lambda_{i}p_{ij})^{y_{j}}}{y_{j}!}.
\]

The conditional entropy of a measurement $y\in Y$ given a distribution
$\lambda\in\Lambda$ of the parameter to be estimated is then given
by \cite{CEM,entropy_def1,entropy_def2} 
\begin{equation}
H\left(y|\lambda\right)=-\sum_{\lambda}P\left(\lambda\right)\left[\sum_{y}P\left(y|\lambda\right)\log P\left(y|\lambda\right)\right]\label{eq:conditional_entropy}
\end{equation}
with $P\left(\lambda\right)$ prior distribution function. The conditional
entropy (\ref{eq:conditional_entropy}) measures the information about
the knowledge of our system, and the realization \mbox{$\hat{\lambda}=\left(\hat{\lambda}_{1},\ldots,\hat{\lambda}_{N}\right)$}
that maximizes (\ref{eq:conditional_entropy}) is the one most likely to be realized. Following
\cite{CEM}, the maximization of $H\left(y|\lambda\right)$ with respect
to the parameter (or the set of parameters) $\lambda$ yields 
\begin{multline}
P\left(y|\hat{\lambda}\right)\left[\lambda_{i}\frac{\partial P}{\partial\lambda_{i}}\log P\left(y|\lambda\right)\right.\\
+\lambda_{i}t_{m}P\left(\lambda_{i}\right)\left(1+\log P\left(y|\lambda\right)\right)\left(\sum_{j=1}^{M}\frac{y_{j}p_{ij}}{\sum_{i=1}^{N}\lambda_{i}p_{ij}}\right)\\
-\left.\lambda_{i}t_{m}\left(1+\log P\left(y|\lambda\right)\right)\sum_{j=1}^{M}p_{ij}\right]_{\lambda_{i}=\hat{\lambda}_{i}}=0\label{eq:entropy_maximization}
\end{multline}

where $\hat{\lambda}_{i}$ are the values of the parameters that maximize
the conditional entropy. Equation \eqref{eq:entropy_maximization}
is verified when the second factor vanishes ($P\left(y|\lambda\right)\neq0$),
leading to 
\begin{multline}
\hat{\lambda}_{i}=\frac{\hat{\lambda}_{i}}{P\left(\hat{\lambda}_{i}\right)\left(1+\log P\left(y|\lambda_{i}\right)\right)t_{m}\sum_{j=1}^{M}p_{ij}}\\
\times\left[\left.\frac{\partial P}{\partial\lambda_{i}}\right|_{\lambda_{i}=\hat{\lambda}_{i}}\log P\left(y|\hat{\lambda}_{i}\right)\right.\\
+\left.P\left(\hat{\lambda}_{i}\right)\left(1+\log P\left(y|\hat{\lambda}_{i}\right)\right)t_{m}\sum_{j=1}^{M}\frac{y_{j}p_{ij}}{\sum_{i=1}^{N}\hat{\lambda}_{i}p_{ij}}\right].\label{eq:solution_entropy_max}
\end{multline}
This equation has no closed solution, thus we need to solve it numerically.
In particular, given the special form of (\ref{eq:solution_entropy_max}),
an iterative solution can be achieved by setting the right-hand-side
as the estimate for the $\left(k+1\right)$-th iterative step of the
left-hand-side: 

\begin{align}
\hat{\lambda}_{i}^{k+1} & =\frac{\hat{\lambda}_{i}^{k}}{P\left(\hat{\lambda}_{i}^{k}\right)\left(1+\log P\left(y|\lambda_{i}^{k}\right)\right)t_{m}\sum_{j=1}^{M}p_{ij}}\label{eq:recursive_full_formula}\\
 & \times\left[\left.\frac{\partial P}{\partial\lambda_{i}}\right|_{\lambda_{i}=\hat{\lambda}_{i}^{k}}\log P\left(y|\hat{\lambda}_{i}^{k}\right)\right.\nonumber \\
 & \left.+P\left(\hat{\lambda}_{i}^{k}\right)\left(1+\log P\left(y|\hat{\lambda}_{i}^{k}\right)\right)t_{m}\sum_{j=1}^{M}\frac{y_{j}p_{ij}}{\sum_{i=1}^{N}\hat{\lambda}_{i}^{k}p_{ij}}\right].\nonumber 
\end{align}
Considering that $P\left(y|\hat{\lambda}\right)$ is a probability
(and thus non-negative) and that in a PET or SPECT system it is much
smaller than 1, we can then use the approximation \mbox{$1+\log P\left(y|\lambda_{i}^{k}\right)\approx\log P\left(y|\lambda_{i}^{k}\right)$}.
This leads to the formula for the estimate of the activity of each
pixel of a waste package: 
\begin{align}
\hat{\lambda}_{i}^{k+1} & =\frac{\hat{\lambda}_{i}^{k}}{P\left(\hat{\lambda}_{i}^{k}\right)t_{m}\sum_{j=1}^{M}p_{ij}}\label{eq:recursive_formula}\\
 & \times\left[\left.\frac{\partial P}{\partial\lambda_{i}}\right|_{\lambda_{i}=\hat{\lambda}_{i}^{k}}+P\left(\hat{\lambda}_{i}^{k}\right)t_{m}\sum_{j=1}^{M}\frac{y_{j}p_{ij}}{\sum_{i=1}^{N}\hat{\lambda}_{i}^{k}p_{ij}}\right].\nonumber 
\end{align}
This means that we can estimate the activity and the position of radioactive
sources if we choose $\lambda_{i}$ to be the activity of pixel $i$.

As it is clear from (\ref{eq:recursive_formula}), the role of the
prior is crucial. What one needs is a function that encompasses all
the prior information about the system and the measurement. In our
case, we have only considered the case of a uniform prior $P\left(\lambda\right)=const.=1$,
meaning that we assume to have no prior knowledge: this choice simplifies
a lot the formula (\ref{eq:recursive_formula}) and reduces our algorithm
to the Maximization of the Likelihood function \cite{entropy_def_1:3,entropy_def_2:4}:
\[
\hat{\lambda}_{i}^{k+1}=\frac{\hat{\lambda}_{i}}{\sum_{j=1}^{M}p_{ij}}\sum_{j=1}^{M}\frac{y_{j}p_{ij}}{\sum_{i=1}^{N}\hat{\lambda}_{i}^{k}p_{ij}}.
\]
Further proposals of priors can be found in \cite{CEM_mesh,CEM_further_priors:7}.

The CEM method provides us a direct way to estimate the position of
radioactive sources and their activity. It has to be pointed out that
the method explained above is an indirect evaluation model for the
activity of a sample (or its portions) since it is necessary to first define
a functional (the conditional entropy) over a parameter space and
then maximize it to find the most likely physical setting.

\section{\label{sec:efficiencies}Efficiencies}

As it has been shown in section \ref{sec:CEM}, a crucial role is played
by the efficiencies $p_{ij}$, i.e. the probability that a decay happening
at cell (pixel) $i$ emits a photon that is detected by detector $j$.
To determine the efficiencies, the following properties of the system are
considered:
\begin{itemize}
\item the sample (composition, geometry, density, nuclear cross section), 
\item the detectors (type, position, geometry, tally), 
\item the relative position of the pixel and the detector and the angle
of view, 
\item the world around the sample and the detectors (the sample carrier,
the air between the sample and the detector, the external shielding
material, and whatever is not directly either the sample or the detector), 
\item the source (the isotopic composition, the emission probability and
energy of the photons), 
\end{itemize}
Given the large number of parameters, this calculation is highly nontrivial
and is practically impossible analytically: it is then necessary to
perform Monte Carlo simulations. In the following, we will first describe
how Monte Carlo N-particle simulations (MCNP) are used to calculate
the energy transported from source inside the sample to one of the
detectors. Then we will show how, starting from the properties of
the detectors (in our case plastic scintillation detectors), we can
transform that into an estimate of the photon count.

The procedure to calculate the efficiencies $p_{ij}$ is then the
following: 
\begin{enumerate}
\item the sample and the detection are modeled in the framework of MCNP; 
\item a standard radioactive source is placed in each cell of the sample; 
\item a Monte Carlo simulation of the propagation of gamma photons through
the system is performed; 
\item the energy deposited on the detectors is converted into a count of
detected photons; 
\item the efficiency is calculated as the ratio of detected intensity and
decay rate.
\end{enumerate}

\subsection{MCNP}

MCNP\textregistered\, is a general-purpose, continuous-energy, generalized-geometry,
time-dependent, Monte Carlo radiation-transport code designed to track
many particle types over broad ranges of energies \cite{mcnp}.

Specific areas of application include, e.g. radiation protection and
dosimetry, radiation shielding, radiography, medical physics, nuclear
criticality safety, detector design and analysis, nuclear oil well
logging, accelerator target design, fission and fusion reactor design,
decontamination and decommissioning. The code treats an arbitrary
three-dimensional configuration of materials in geometric cells bounded
by first- and second-degree surfaces.

The transport is based on point-wise cross-section data. For photons,
the code accounts for incoherent and coherent scattering, the possibility
of fluorescent emission after photoelectric absorption, absorption
in pair production with local emission of annihilation radiation,
and bremsstrahlung.

The code provides easy-to-use and very versatile general sources and
flexible tallies. In the specific case of the calculation of efficiencies
for detectors in a Large Clearance Monitor the sources are modeled
by a isotropic point source located in the center of each pixel. The
energy distribution of the source photons reflects the decay gamma
energies for the respective radio nuclide. The code then transports
the particles through the geometry and a pulse-height tally which
records the energy deposited in the detector volume by each primary
particle and its secondary particles. The spectrum of the deposited
energies is calculated on a very fine resolution (\textasciitilde 3000
bins) between 0 eV and 3 keV.

\subsection{SimPS}

SimPS (Simulated Plastic Scintillator) is a software for plastic scintillation
detectors to calculate detection probabilities of photons with specific
energies which interact with the detector. Moreover, it takes into
account the emission probability, i.e. the number of photons emitted
per decay of the radioactive source. In order to do so, a detector
characterization and calibration needs to be performed by estimating
the energy dependence of the detection probability. This probability
is mainly influenced by the discriminator voltage of the detector
and can be modeled by means of a cumulative distribution function
of a Pareto-distribution (\ref{eq:pareto}) whose parameters $x_{m}$
and $\alpha$ are estimated by fitting calibration measurements 
\begin{equation}
F(x)=\begin{cases}
1-\frac{x_{m}^{\alpha}}{x}, & x\geq x_{m}\\
0, & x<x_{m}
\end{cases}.\label{eq:pareto}
\end{equation}

Using this detector characterization and the energy deposition spectrum
calculated by MCNP for each detector-pixel pair, it is possible to
estimate the efficiencies $p_{ij}$ used in the CEM algorithm.

\section{\label{sec:cem_fma}Application of the CEM to clearance measurements}

\subsection{Compliance with the ISO 11929}

As it has been already pointed out, the analysis of a potentially
contaminated sample is subject to the compliance with the international
guidelines set by ISO and adopted by the national governments \cite{DIN_ISO_11929:8}.
As it has been reported in section \ref{subsec:DIN_ISO}, the idea behind
the standard is that not only the best estimate of the activity is
needed to make a decision about a sample, but also its distribution
has to be fully characterized by calculating all the statistical characteristic
values, i.e. the coverage interval, the detection limit and the decision
threshold. In order to achieve this for each measured sample, we need
an evaluation model to estimate the activity and its uncertainty once
we have performed the measurements as it has been already explained.

It is then crucial to show if and how the method described in the
previous section is able to estimate activity values and their uncertainty.
The norm prescribes to use an algebraic model for the answer to this
question, and thus it might seem that the CEM method is not compliant
with the ISO 11929:2019. In fact, the CEM method is not only an imaging
technique, but also a way to estimate the activity (and the respective
uncertainty) of any portion of a sample, making it suitable to our
purpose.

Looking back at the general formulation of the CEM method, we can
fix the general parameter $\lambda$ to be the activity of a cell
of the sample and this allows the CEM to produce an estimate of the
activity of each cell: this is a good result, but still incomplete
as the CEM itself provides no direct method to characterize the distribution
of the activity. We will address this problem in the following section \ref{sec:characteristic_limits}; for now it is enough to know that the CEM algorithm can be opportunely integrated to fully characterize the activity distribution of each cell of a measured sample. This puts
us in the condition to be compliant with the ISO 11929:2019,
since we have both a model of evaluation of the activity (resp. activities
of the single cells or portions of the sample) and an estimate of the uncertainty
(resp. uncertainty for each cell or portions of the sample). In the following section \ref{subsec:net_to_gross}
we will show our proposal to calculate all the characteristic values
of the activity distribution for single cells and for whatever portion
of the sample, showing also that this improves a lot the limits of
the coverage interval of the activity of the whole sample.

\subsection{\label{subsec:net_to_gross}Application of the CEM to the clearance
of waste packages}

The general CEM method presented in section \ref{sec:CEM} has a direct and
easy application for PET imaging systems. Despite the application
to SPECT systems is straightforward as well, there are some technical
difficulties that make it tricky. In fact, in a PET system, given
the nature of the emission process, the device measures coincidence
count rates and thus can directly measure the net count rate of pair
annihilations. On the other hand, in a SPECT system (as it is in our
case), the detectors measure also the background radiation (due to
natural radioactive sources) and thus one can either subtract the
background from the gross count rate, or slightly change the formula
(\ref{eq:recursive_formula}) in order to use the gross count rate
instead. Using the gross count rate is of help because in our case the count due to the sample can be similar or even lower than the count of background photons. Moreover, using the gross count rate will make also the calculation of the uncertainties easier, as it will be shown in the following. The most straightforward way of
circumventing the problem is to substitute the mean count (\ref{eq:mean_value_poissonian})
with the corresponding expression of the gross number of emitted photons
according to the evaluation model (\ref{eq:eval_model}) of the ISO
11929:2019 
\begin{equation}
\mu_{j}=t_{m}\sum_{i=1}^{N}\lambda_{i}p_{ij}\longrightarrow t_{m}\left(x_{3}r_{0,j}-x_{42}+\sum_{i=1}^{N}\lambda_{i}p_{ij}\right).\label{eq:substitution_mean_count}
\end{equation}
This modifies the formula (\ref{eq:solution_entropy_max}) leading
to
\begin{multline*}
\hat{\lambda}_{i}=\frac{\hat{\lambda}_{i}}{P\left(\hat{\lambda}_{i}\right)\left(1+\log P\left(y|\lambda_{i}\right)\right)t_{m}\sum_{j=1}^{M}p_{ij}}\\
\left[\left.\frac{\partial P}{\partial\lambda_{i}}\right|_{\lambda_{i}=\hat{\lambda}_{i}}\log P\left(y|\hat{\lambda}_{i}\right)+\right.\\
+P\left(\hat{\lambda}_{i}\right)\left(1+\log P\left(y|\hat{\lambda}_{i}\right)\right)t_{m}\\
\times\left.\sum_{j=1}^{M}\frac{y_{j}p_{ij}}{r_{0,j}x_{3}-x_{42}+\sum_{i=1}^{N}\hat{\lambda}_{i}p_{ij}}\right]
\end{multline*}
where $y=\left(y_{1},\ldots,y_{N}\right)$ are now the gross count
rates of the $N$ detectors. It has to be pointed out that the substitution
(\ref{eq:substitution_mean_count}) is also needed to calculate the
characteristic statistical limits of the activity distribution as
it will be shown in the next section. The recursive formula (\ref{eq:recursive_formula})
for the estimate of the activity becomes then
\begin{multline}
\hat{\lambda}_{i}^{k+1}=\frac{\hat{\lambda}_{i}^{k}}{P\left(\hat{\lambda}_{i}^{k}\right)t_{m}\sum_{j=1}^{M}p_{ij}}\left[\left.\frac{\partial P}{\partial\lambda_{i}}\right|_{\lambda_{i}=\hat{\lambda}_{i}^{k}}\right.\\
\left.+P\left(\hat{\lambda}_{i}^{k}\right)t_{m}\sum_{j=1}^{M}\frac{y_{j}p_{ij}}{r_{0,j}x_{3}-x_{42}+\sum_{i=1}^{N}\hat{\lambda}_{i}^{k}p_{ij}}\right].\label{eq:recursive_formula_gross}
\end{multline}
As a direct consequence, also the maximum likelihood formula (i.e.
with constant prior) is modified: 
\[
\hat{\lambda}_{i}^{k+1}=\frac{\hat{\lambda}_{i}^{k}}{\sum_{j=1}^{M}p_{ij}}\sum_{j=1}^{M}\frac{y_{j}p_{ij}}{r_{0,j}x_{3}-x_{42}+\sum_{i=1}^{N}\hat{\lambda}_{i}^{k}p_{ij}}.
\]

\section{\label{sec:characteristic_limits}Calculation of the characteristic
limits}

The first step to calculate the characteristic statistical limits
of the activity distribution is to estimate the uncertainty of the
activities calculated by means of the CEM algorithm. As we have already pointed out, the CEM algorithm itself
does not provide a direct way to estimate the uncertainty, but this
problem can be easily circumvented by propagating the uncertainty
of the measurement and of any other input parameter (according
to their statistical distribution). The measured values $y_{i}$ are count rates of photons,
and thus they are Poisson distributed.
The idea is then to generate $P$ Poisson random sets of data \mbox{$y^{\eta}=\left(y_{1}^{\eta},\ldots,y_{M}^{\eta}\right)$}, \mbox{$\eta=1,\ldots,P$} (or repeating the measurement $P$ times) and perform
the CEM calculation for each of this randomly generated (or measured)
set of data. This will provide $P$ different activity distributions
$\lambda^{\eta}=\left(\lambda_{1}^{\eta},\ldots,\lambda_{n}^{\eta}\right)$
and then we can use the obtained values to calculate the activity
and the uncertainty for each portion of the sample (in the most simple
case for each single cell) as the arithmetic mean and the standard
deviation of the values produced by the CEM algorithm: 
\begin{align}
\lambda_{i}= & \frac{1}{P}\sum_{\eta=1}^{P}\lambda_{i}^{\eta}\label{eq:mean}\\
u_{i}^{2}\left(\lambda_{i}\right)= & \frac{1}{P-1}\sum_{\eta=1}^{P}\left(\lambda_{i}^{\eta}-\lambda_{i}\right)^{2}.\label{eq:stdev}
\end{align}
This procedure can be repeated if we have any other known source of
uncertainty for our count rates. In particular, looking at (\ref{eq:recursive_formula_gross}),
it is necessary to propagate also the uncertainty of the background
$r_{0}$, of the efficiencies $p_{ij}$ (that are normally distributed),
of the factor $x_{3}$ and of the shielding $x_{42}$.

Each cell will then have a distribution of estimated activities
\mbox{$\left(\lambda_{i}^{1},\ldots,\lambda_{i}^{P}\right)$} that we can
sort such that \mbox{$\lambda_{i}^{\eta}\leq\lambda_{i}^{\eta+1}$}. Then we
can assign to each value $\lambda_{i}^{\eta}$ a cumulative probability
$\eta/P$ and build a discrete cumulative distribution $F\left(\lambda_{i}|y^{\eta}\right)$.

We now have all the ingredients to fully characterize the statistical
activity distributions of each portion of the sample. This can be
done by following the guidelines of the ISO 11929:2019, i.e. by
estimating the uncertainty for specific values of the assumed true
value. We need then to calculate the outcome of a measurement $\tilde{y}_{j}$
when the activities $\lambda_{i}$ assume some special values $\tilde{\lambda}_{i}$.
This can be done following (\ref{eq:substitution_mean_count}): 
\[
\tilde{y}_{j}=x_{3}r_{0,j}-x_{42}+\sum_{i=1}^{N}\tilde{\lambda}_{i}p_{ij}.
\]

Then the $\tilde{y}_{j}$ have to be varied around the mean and given
as input to the CEM formula (\ref{eq:recursive_formula_gross}) to
estimate the characteristic limits pixel by pixel 
\begin{align}
\hat{\lambda}_{i}^{k+1} & =\frac{\hat{\lambda}_{i}^{k}}{P\left(\hat{\lambda}_{i}^{k}\right)t_{m}\sum_{j=1}^{M}p_{ij}}\left[\left.\frac{\partial P}{\partial\lambda_{i}}\right|_{\lambda_{i}=\hat{\lambda}_{i}^{k}}\right.\label{eq:recusive_formula_char_lim}\\
 & \left.+P\left(\hat{\lambda}_{i}^{k}\right)t_{m}\sum_{j=1}^{M}\frac{\left(x_{3}r_{0,j}-x_{42}+\sum_{i=1}^{N}\tilde{\lambda}_{i}p_{ij}\right)p_{ij}}{r_{0}x_{3}-x_{42}+\sum_{i=1}^{N}\hat{\lambda}_{i}^{k}p_{ij}}\right].\nonumber 
\end{align}
The mean and the standard deviation are then calculated using (\ref{eq:mean})
and (\ref{eq:stdev}).

Formula \eqref{eq:recusive_formula_char_lim} can be seen as the bridge
between the CEM algorithm (i.e. the evaluation model for all the cells) and the statistical characteristic limits
(for single cells or blocks). In the following we will indeed address
the problem not only for single cells, but also for blocks, that can
be reduced to single-cell-blocks. To this end we can split the sample itself into
two parts $B_{1}$ (the part of interest, of which we want to calculate
a characteristic limit) and $B_{2}$ (the rest), such that we can
split the sum $\sum_{i=1}^{N}\lambda_{i}p_{ij}$ in (\ref{eq:recusive_formula_char_lim})
into two parts as follows 
\[
\sum_{i=1}^{N}\lambda_{i}p_{ij}=\sum_{i\in B_{1}}\lambda_{i}p_{ij}+\sum_{i\in B_{2}}\lambda_{i}p_{ij}
\]

where a block can consist even of one single cell or all the cells.
In order to keep the pattern of the estimated activity distribution
fixed in $B_{1}$ and $B_{2}$ respectively while varying the assumed
true value we define the specific efficiency for the two blocks 
\begin{eqnarray*}
p_{B_{1},j}=\frac{\sum_{i\in B_{1}}\lambda_{i}p_{ij}}{\sum_{i\in B_{1}}\lambda_{i}} & ; & p_{B_{2},j}=\frac{\sum_{i\in B_{2}}\lambda_{i}p_{ij}}{\sum_{i\in B_{2}}\lambda_{i}}.
\end{eqnarray*}
With these constrains the iterative CEM formula for the activity in
$B_{1}$ can be written as follows 
\begin{multline}
\hat{\lambda}_{B_{1}}^{k+1}=\frac{\hat{\lambda}_{B_{1}}^{k}}{P\left(\hat{\lambda}_{B_{1}}^{k}\right)t_{m}\sum_{j=1}^{M}p_{B_{1},j}}\\
\times\left[\left.\frac{\partial P}{\partial\lambda_{B_{1}}}\right|_{\lambda_{B_{1}}=\hat{\lambda}_{B_{1}}^{k}}+P\left(\hat{\lambda}_{B_{1}}^{k}\right)t_{m}\right.\\
\left.\times\sum_{j=1}^{M}\frac{\left(x_{3}r_{0,j}-x_{42}+\tilde{\lambda}_{B_{1}}p_{B_{1},j}+\tilde{\lambda}_{B_{2}}p_{B_{2},j}\right)p_{B_{1},j}}{r_{0,j}x_{3}-x_{42}+\tilde{\lambda}_{B_{2}}p_{B_{2},j}+\hat{\lambda}_{B_{1}}^{k}p_{B_{1},j}}\right].\label{eq:char_limit_block_1}
\end{multline}
Assuming a true value for $\tilde{\lambda}_{B_{1}}$, we can then
sample the input parameters and estimate the characteristic values. It has to be remarked
that the numerator and the denominator of (\ref{eq:char_limit_block_1}) have to be sampled separately
as they represent measurements of the background that happen respectively with (numerator)
and without (denominator) the sample (i.e. the block of interest) in the chamber.

\subsection{Decision threshold}

Following the prescriptions of the ISO 11929:2019, if we want
to calculate the decision threshold of the activity of a sample, we
need to estimate the uncertainty of a vanishing true value of the
activity. The decision threshold can be defined as the minimum value
that can be distinguished from the background with a given confidence
and in this case the definition of background itself is crucial. In
order to calculate the decision threshold for a block of pixels (or
eventually for a single pixel), we need to consider the radiation
emitted by the rest of the sample as part of the background.

In order to calculate the decision threshold for $B_{1}$, we need
to set its total activity $\tilde{\lambda}_{B_{1}}=0$ while the activities
of the cells in $B_{2}$ are left unchanged (i.e. the value estimated
by the CEM) and are taken into account as part of the background radiation.
This can be obtained only by setting all the $\tilde{\lambda}_{i}$
in $B_{1}$ to zero (since activities are non negative), leading to
the following iterative formula for the decision threshold $\lambda_{B_{1}}^{*}$
of block $B_{1}$ that follows from (\ref{eq:char_limit_block_1})
\begin{multline*}
\hat{\lambda}_{B_{1}}^{k+1}=\frac{\hat{\lambda}_{B_{1}}^{k}}{P\left(\hat{\lambda}_{B_{1}}^{k}\right)t_{m}\sum_{j=1}^{M}p_{B_{1},j}}\left[\left.\frac{\partial P}{\partial\lambda_{B_{1}}}\right|_{\lambda_{B_{1}}=\hat{\lambda}_{B_{1}}^{k}}\right.\\
\left.+P\left(\hat{\lambda}_{B_{1}}^{k}\right)t_{m}\sum_{j=1}^{M}\frac{\left(x_{3}r_{0,j}-x_{42}+\tilde{\lambda}_{B_{2}}p_{B_{2},j}\right)p_{B_{1},j}}{r_{0,j}x_{3}-x_{42}+\tilde{\lambda}_{B_{2}}p_{B_{2},j}+\hat{\lambda}_{B_{1}}^{k}p_{B_{1},j}}\right].
\end{multline*}
By sampling the input parameters of this formula around the fixed
true values ($r_{0}$,$x_{3}$,$x_{42}$,$\tilde{\lambda}_{B_{2}},\text{\ensuremath{p_{B_{1},j}}},p_{B_{2},j}$),
we get many values of $\lambda_{B_{1}}^{\eta}$ that are distributed
according to $F\left(\tilde{\lambda}_{B_{1}}=0|y^{\eta}\right)$ as
described in the previous paragraph. The $\left(1-\alpha\right)$-quantile
(typically with $\alpha=0.05$) of this distribution is the decision
thresholt $\lambda_{B_{1}}^{*}$.

\subsection{Detection limit}

The detection limit $\lambda_{B_{1}}^{\#}$ of $B_{1}$ is the smallest
true value of the activity in the block, for which the probability
of a false negative (as it is explained in section \ref{sec:appendix_characteristic_limits})
does not exceed the specified probability $\beta$ (typically 0.05).

By applying the iterative formula (\ref{eq:char_limit_block_1}),
the detection limit $\lambda_{B_{1}}^{\#}$ is the value of $\tilde{\lambda}_{B_{1}}$
for which the $\beta$-quantile of the distribution function $F\left(\tilde{\lambda}_{B_{1}}|y^{\eta}\right)$
is equal to the decision threshold. This value can to be calculated
iteratively by applying root-finding methods.

\subsection{\label{subsec:coverage_interval}Limits of the coverage interval}

The limits of the coverage interval ($\lambda_{B_{1}}^{\triangleleft}$,
$\lambda_{B_{1}}^{\triangleright}$) are defined in such a way that
the coverage interval contains the true value of the measurand within
the specified probability $1-\gamma$ (typically with $\gamma=0.05$).
In the following we use the definition of the probabilistic symmetric
coverage interval which defines the limits of the coverage interval
as the $\gamma/2$-quantile ($\lambda_{B_{1}}^{\triangleleft}$) and
$\left(1-\gamma/2\right)$-quantile ($\lambda_{B_{1}}^{\triangleright}$)
of the distribution function $F\left(\lambda_{B_{1}}|y^{\eta}\right)$.

\section{\label{sec:examples}Examples}

In the following we will show the accuracy of the proposed method
by simulating measurements with MCNP and reconstructing the sources
by means of the CEM method. The two examples that follow will summarize
all the things that have been discussed so far and will show that
the method presented is safe and reliable enough to handle waste packages
from nuclear facilities.

\begin{figure}
\begin{centering}
\includegraphics[scale=0.35]{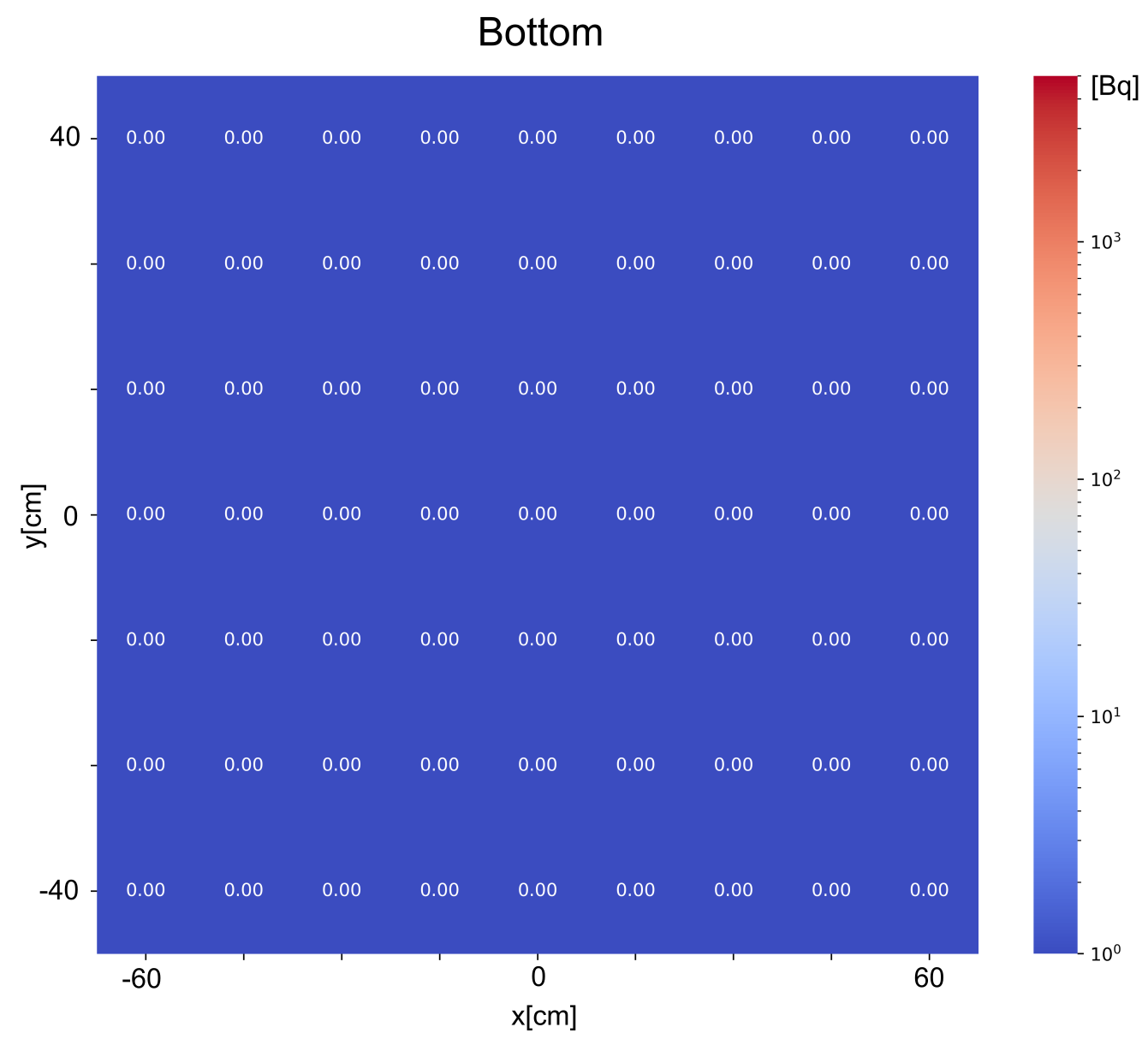}
\par\end{centering}
\begin{centering}
\includegraphics[scale=0.35]{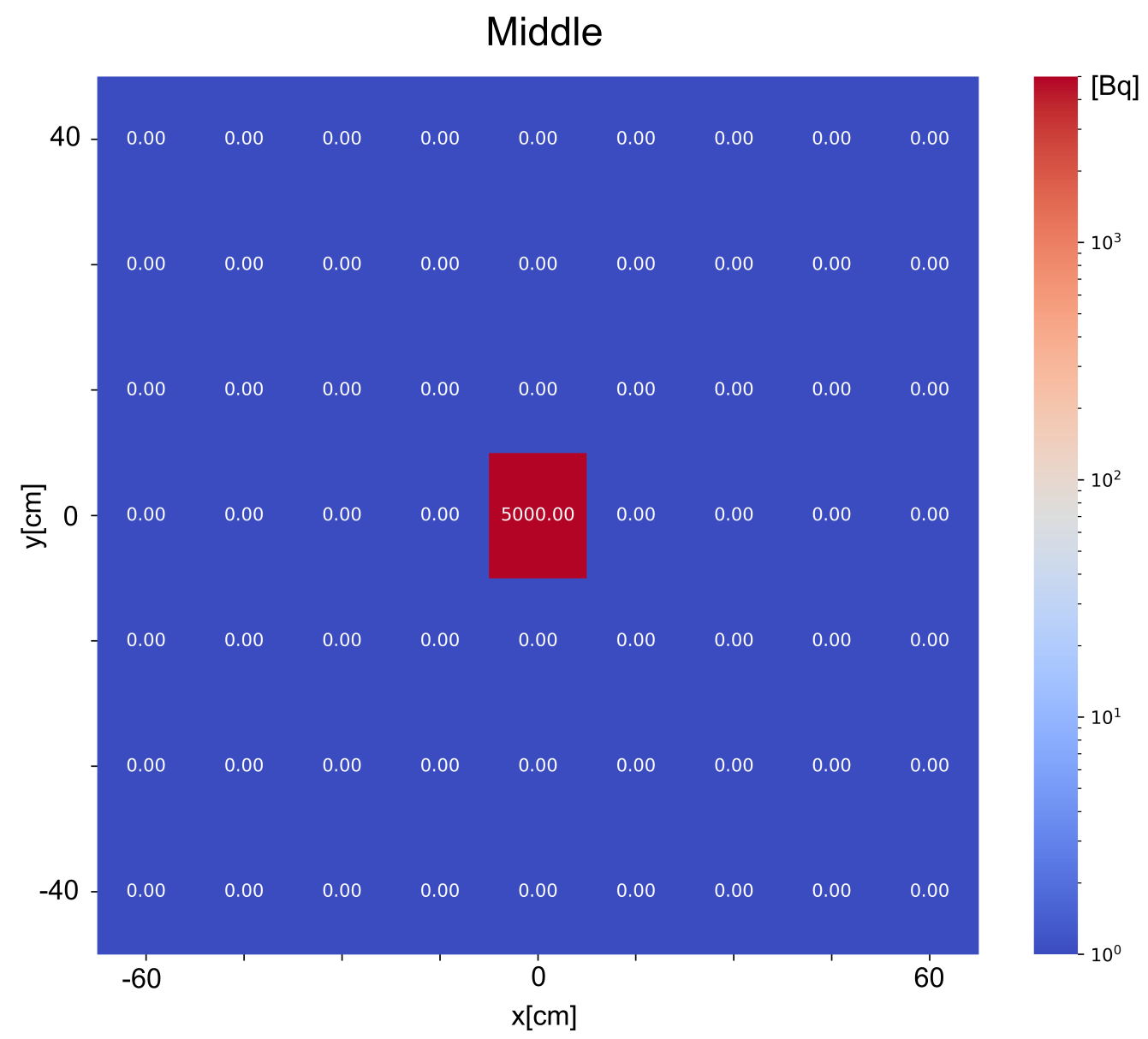}
\par\end{centering}
\begin{centering}
\includegraphics[scale=0.35]{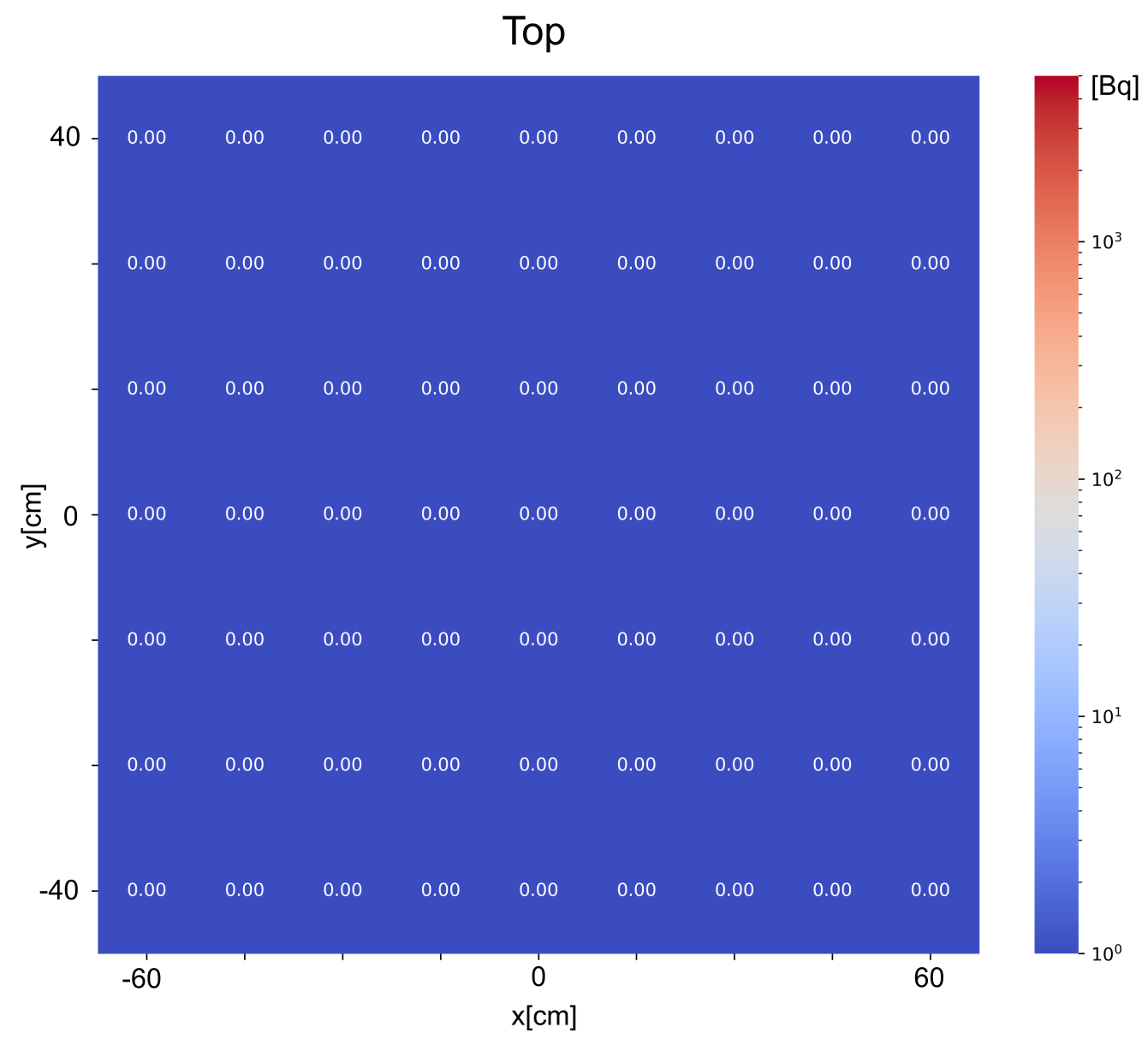} 
\par\end{centering}
\caption{\label{fig:1HS_original}Simulated sample n.1. Uniform iron block
of density 1.681 $g/cm^{3}$ with one simulated $^{60}Co$ source
in the middle of activity 5 $kBq$. The three plots represent the
three layers of the sample from bottom to top.}
\end{figure}

\begin{figure}
\begin{centering}
\includegraphics[scale=0.35]{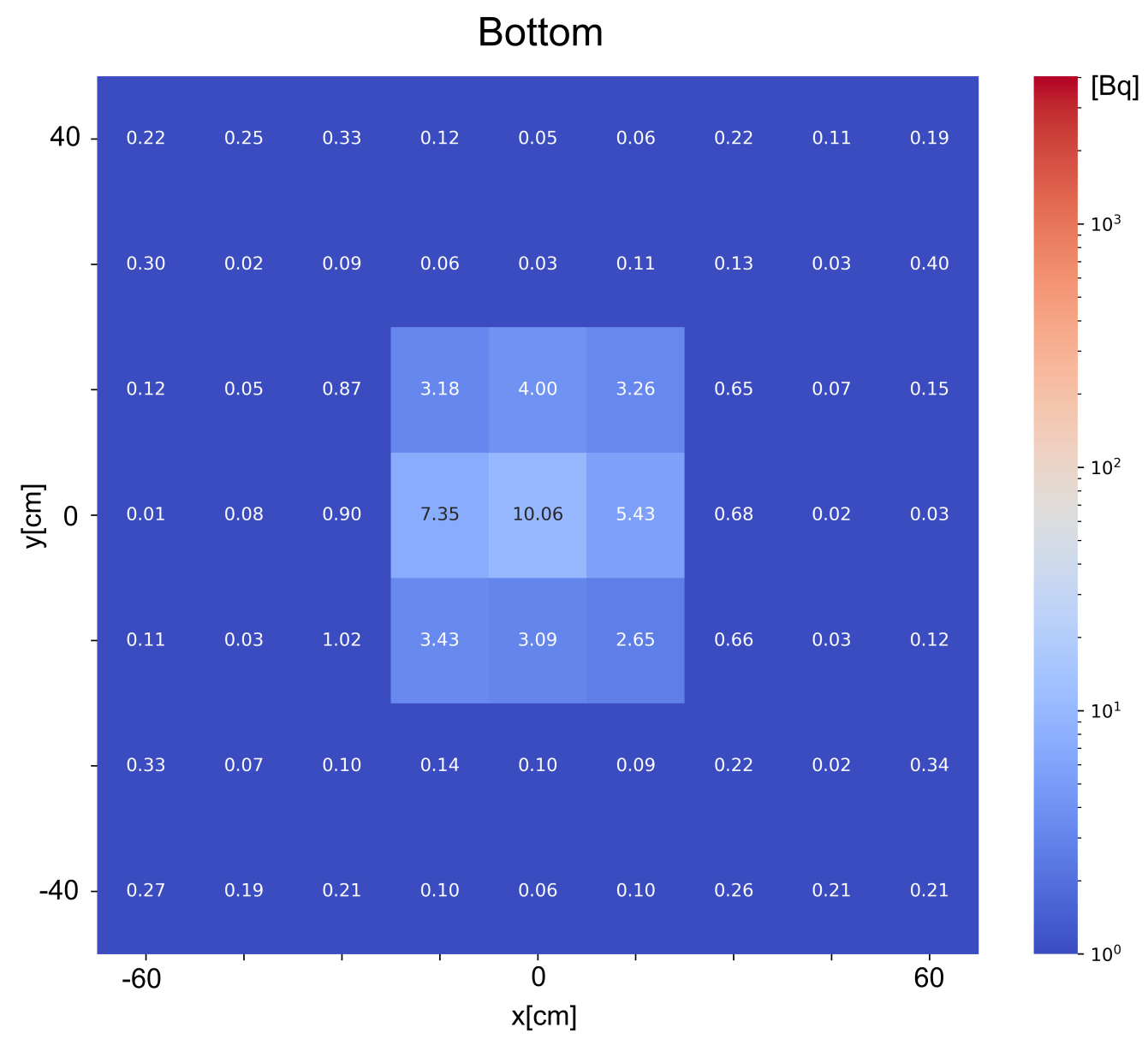}
\par\end{centering}
\begin{centering}
\includegraphics[scale=0.35]{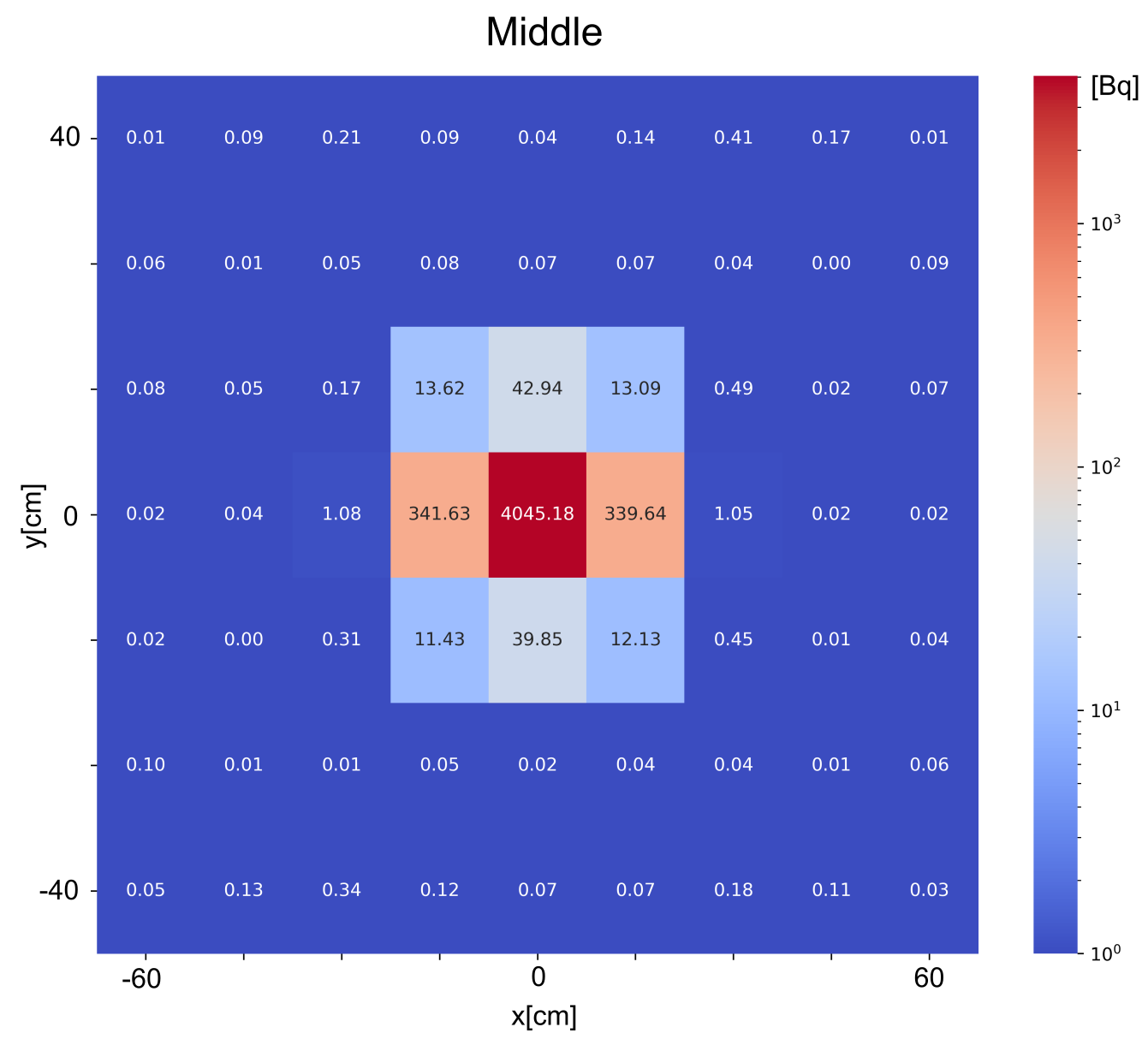} 
\par\end{centering}
\begin{centering}
\includegraphics[scale=0.35]{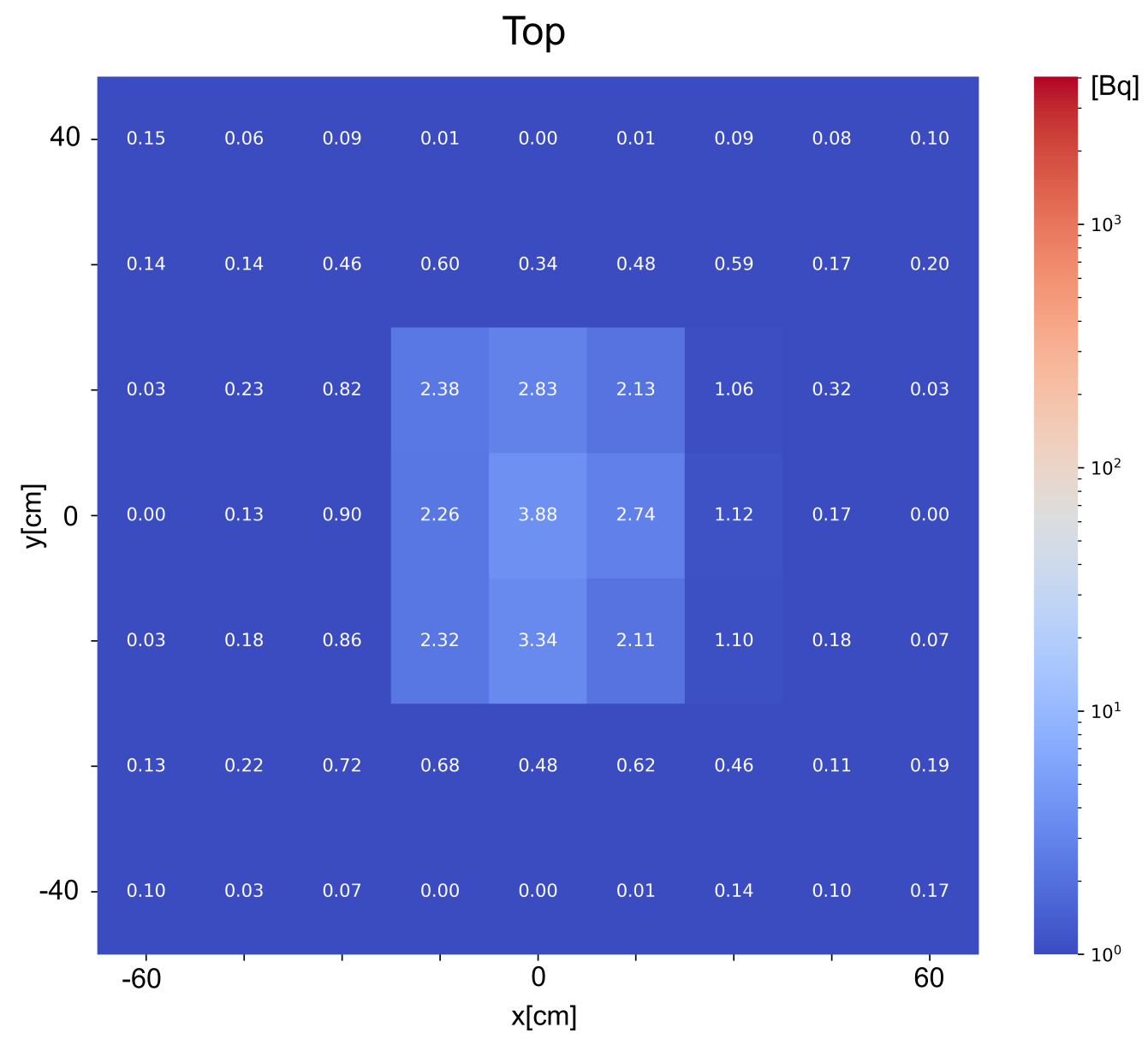} 
\par\end{centering}
\caption{\label{fig:1HS_reconstructed}Simulated sample n.1. This figure shows
the CEM-reconstructed activity distribution of the sample shown in
Fig. \ref{fig:1HS_original}}
\end{figure}

The first simulated sample is an iron block of size 120 x 80 x 23.5
cm (x, y and z dimension) and density 1.681 $g/cm^{3}$ carried by a
steel sample-carrier-box and an aluminum pallet. A point source of
$^{60}Co$ of activity 5 $kBq$ is placed right in the middle of the
sample as it is shown in Fig. \ref{fig:1HS_original} while the CEM-reconstructed
distribution is in Fig. \ref{fig:1HS_reconstructed}. For the reconstruction,
we have sampled the input parameters 10000 times around the measured (or in this case simulated) values
according to their statistical distribution. For each sampled set of data, the recursive
CEM formula \eqref{eq:recursive_formula_gross} has been used with
$k=10^{4}$ to guarantee that the calculations have satisfactorily
converged. The activity for each pixel is then calculated
as the arithmetic mean of all the 10000 runs of the algorithm. The
characteristic values have been calculated according to the prescriptions
layed out in the previous section. For the whole sample, the total
activity is $\lambda_{tot}=4890\:Bq$ with a standard deviation of
$\sigma_{tot}=116\:Bq$. The lower and upper limit of the 95\% symmetric
coverage interval interval are $\lambda_{tot}^{\triangleleft}=4695\:Bq$
and $\lambda_{tot}^{\triangleright}=5077\:Bq$. The decision threshold
is $\lambda_{tot}^{\star}=60\:Bq$ and the detection limit $\lambda_{tot}^{\#}=124\:Bq$.
The integral efficiency of the whole detection system is 21.2\%
for that activity distribution. The characteristic values of this
sample are summarized in Table \ref{tab:table_1HS}.

\begin{table}
\begin{centering}
\begin{tabular}{|c|c|}
\cline{2-2} 
\multicolumn{1}{c|}{} & Total\tabularnewline
\hline 
Activity (Bq)  & 4890\tabularnewline
\hline 
Uncertainty (Bq)  & 116\tabularnewline
\hline 
Lower limit of coverage (Bq)  & 4695\tabularnewline
\hline 
Upper limit of coverage (Bq)  & 5077\tabularnewline
\hline 
Decision threshold (Bq)  & 60\tabularnewline
\hline 
Detection limit (Bq)  & 124\tabularnewline
\hline 
Efficiency (\%)  & 21.2\tabularnewline
\hline 
\end{tabular}
\par\end{centering}
\caption{\label{tab:table_1HS}Table with all the characteristic values of
the simulated sample n. 1 (activity distribution of Fig. \ref{fig:1HS_original})}
\end{table}

\begin{figure}
\begin{centering}
\includegraphics[scale=0.35]{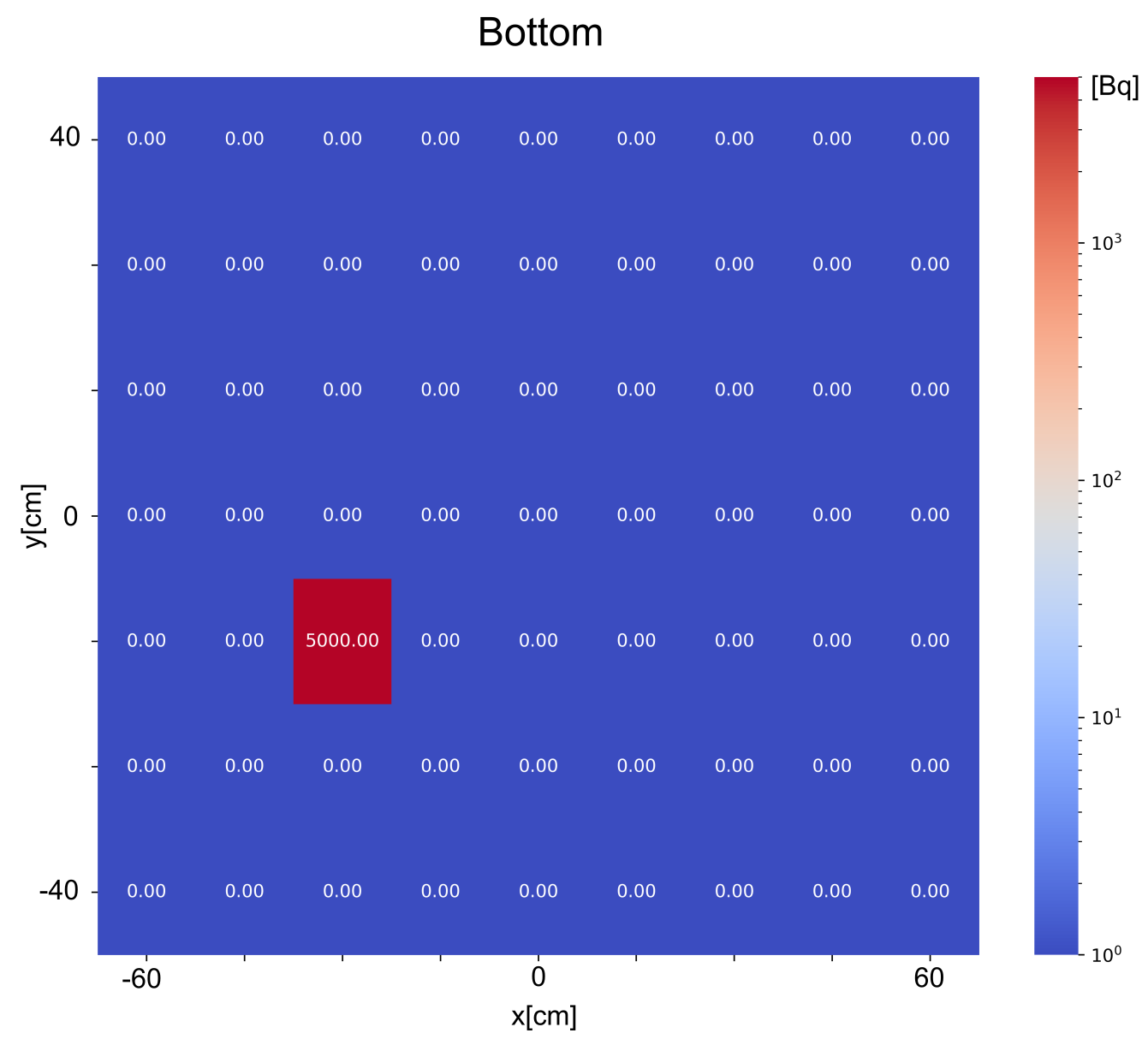}
\par\end{centering}
\begin{centering}
\includegraphics[scale=0.35]{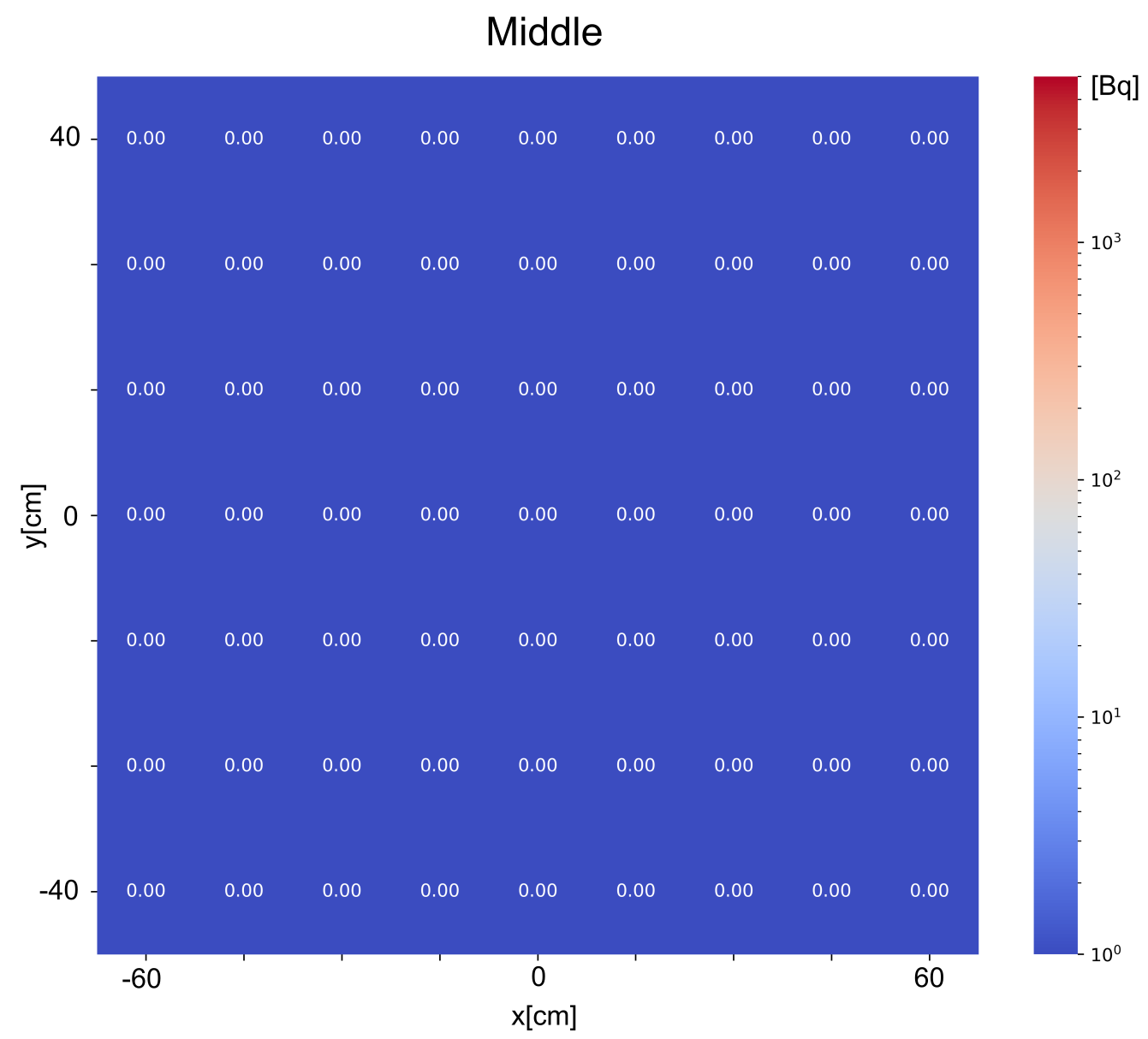} 
\par\end{centering}
\begin{centering}
\includegraphics[scale=0.35]{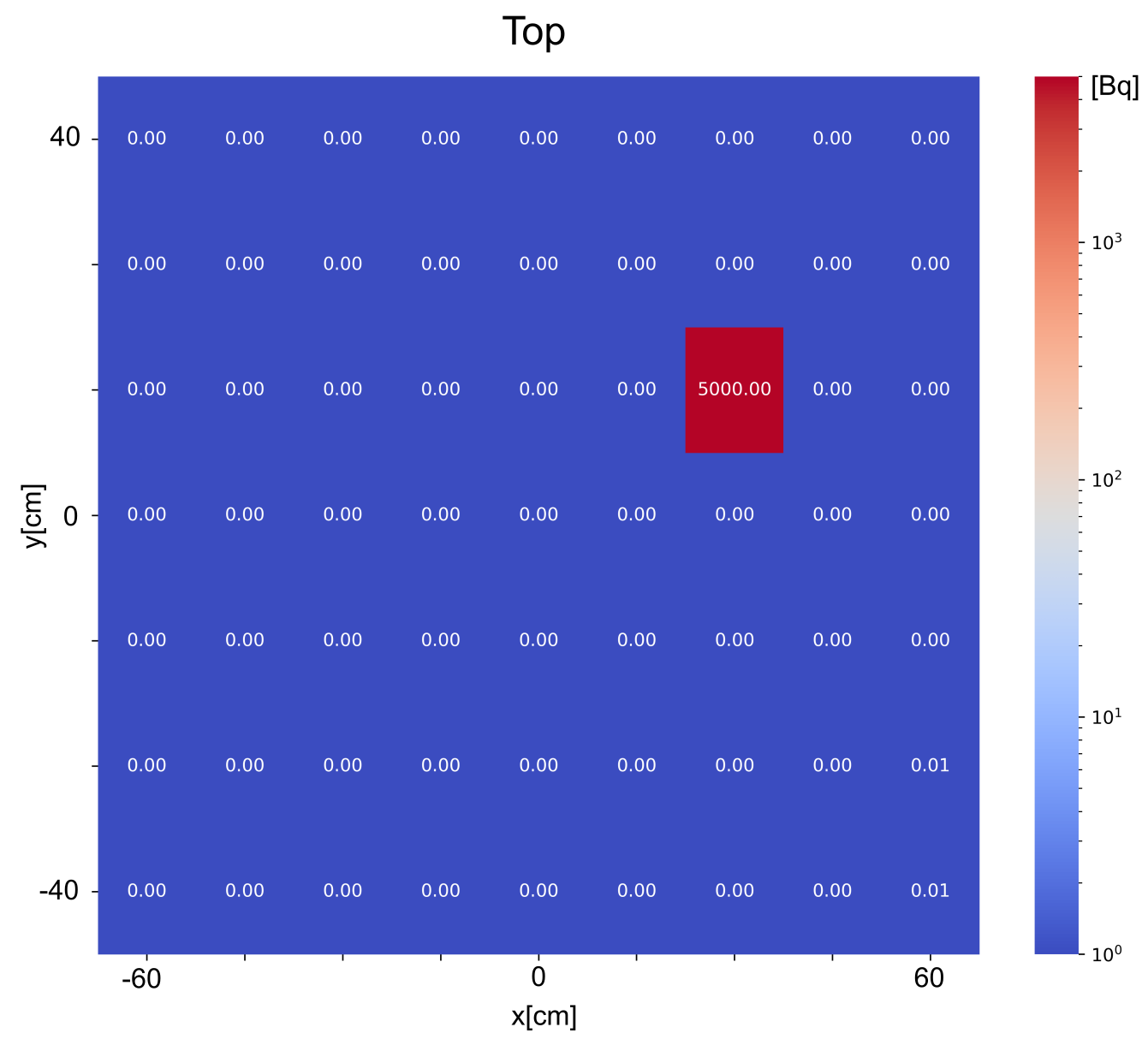} 
\par\end{centering}
\caption{\label{fig:2HS_original}Simulated sample n.2. Uniform iron block
of density 1.681 $g/cm^{3}$ with two simulated $^{60}Co$ sources
of activity 5 $kBq$ each. The three plots represent the three layers
of the sample from bottom to top.}
\end{figure}

\begin{figure}
\begin{centering}
\includegraphics[scale=0.35]{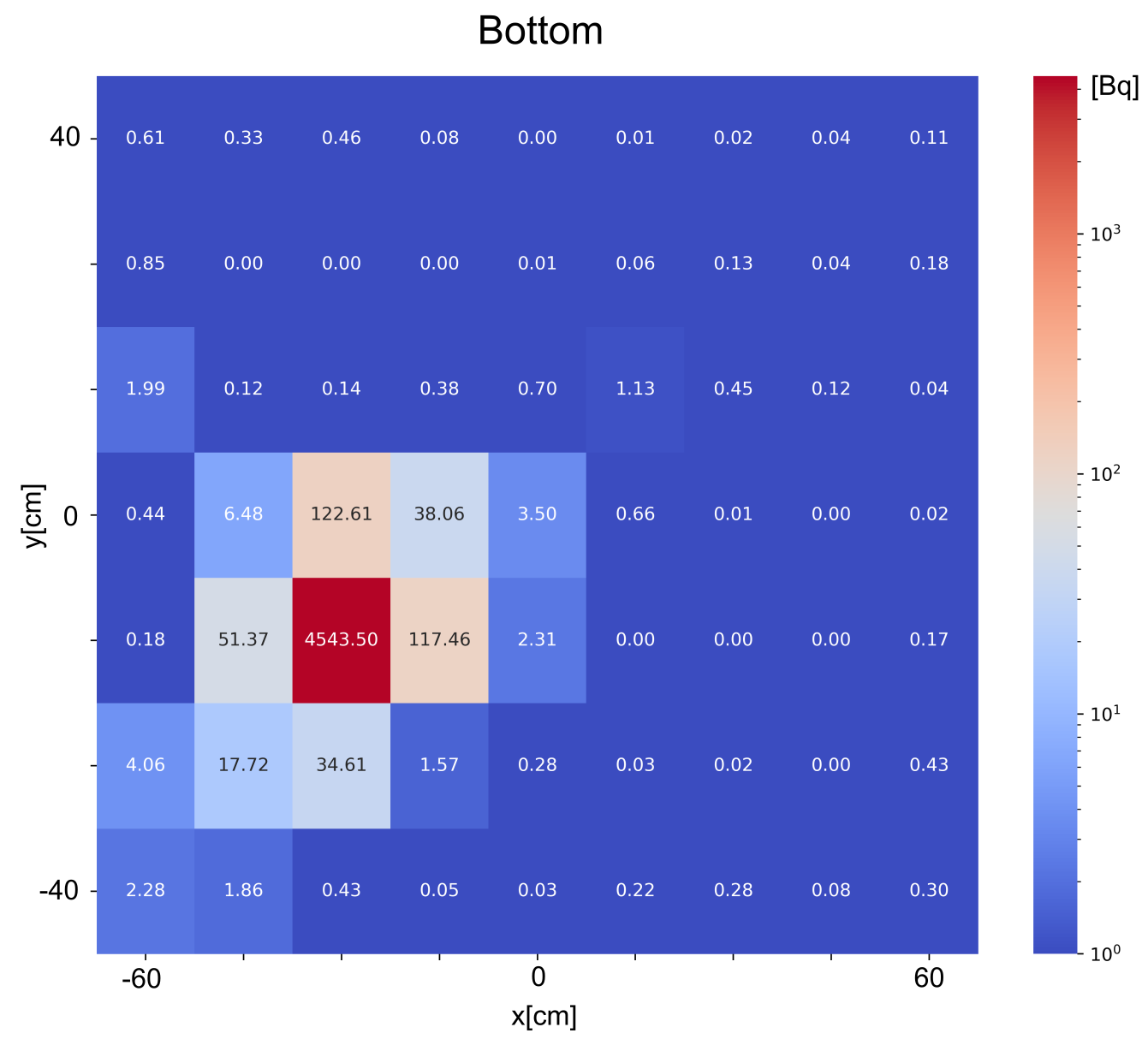}
\par\end{centering}
\begin{centering}
\includegraphics[scale=0.35]{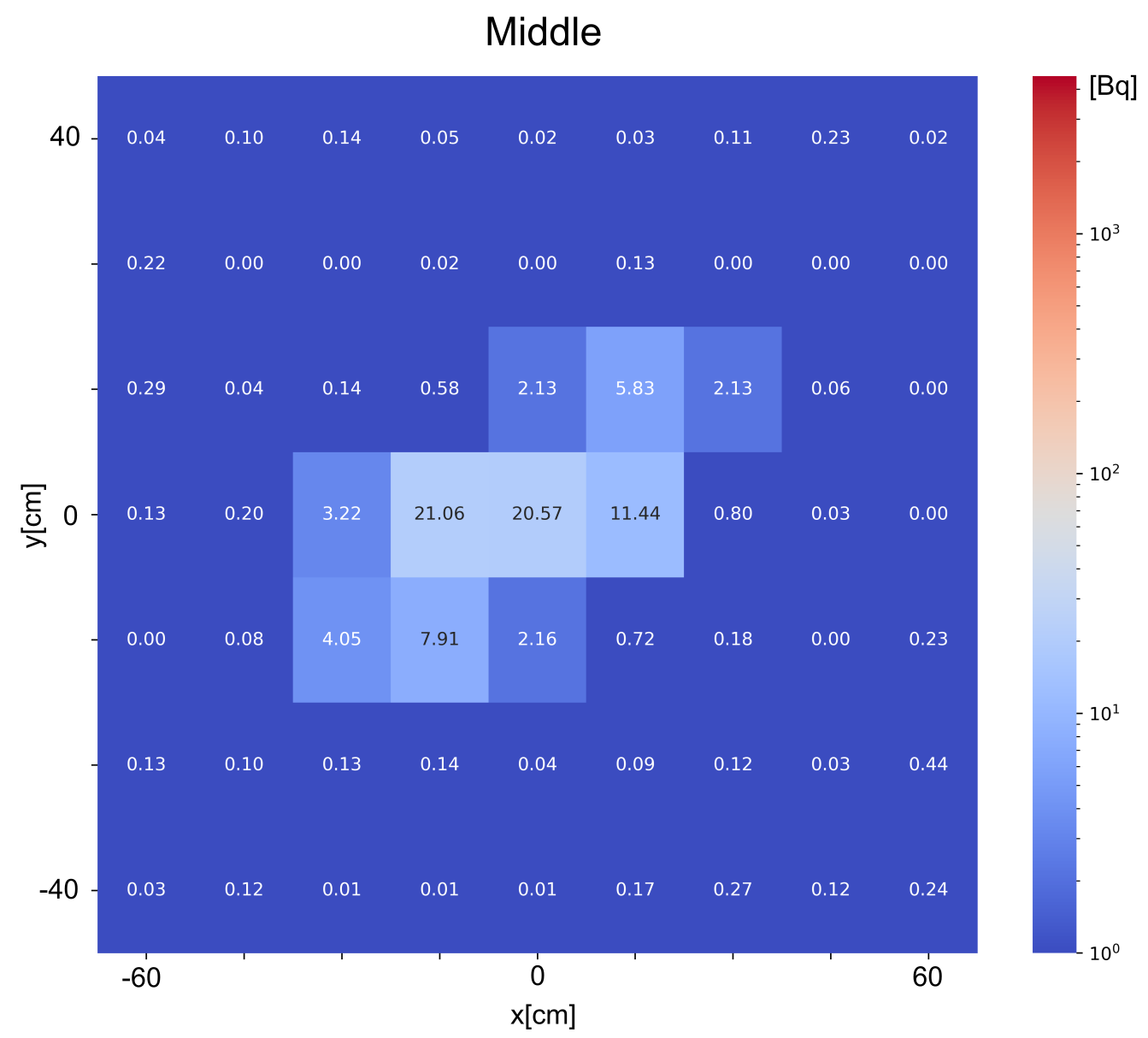} 
\par\end{centering}
\begin{centering}
\includegraphics[scale=0.35]{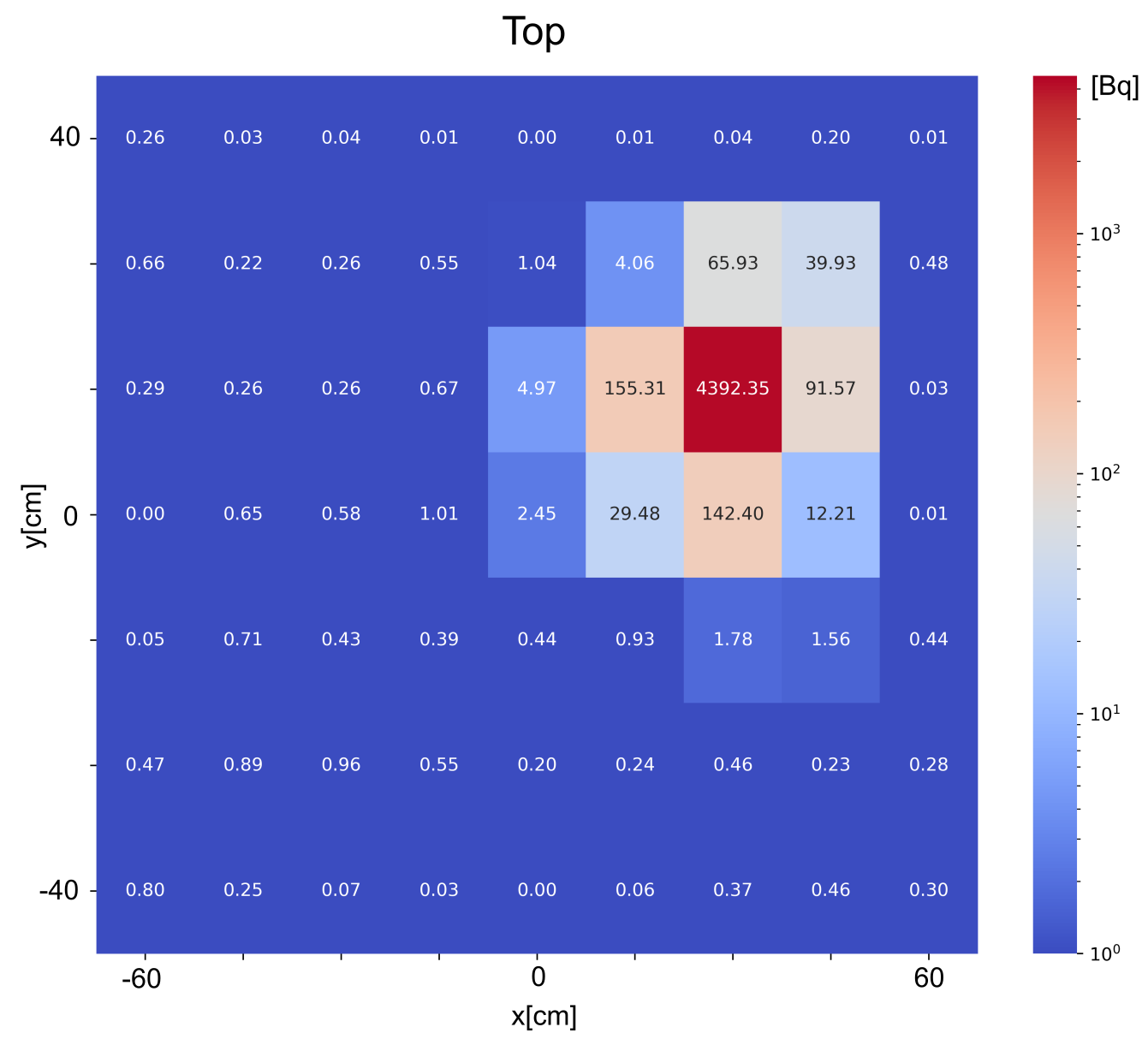} 
\par\end{centering}
\caption{\label{fig:2HS_reconstructed}Simulated sample n.2. This figure shows
the CEM-reconstructed activity distribution of the sample shown in
Fig. \ref{fig:2HS_original}}
\end{figure}

The second simulated sample consists of the same iron block, with
two $^{60}Co$ point sources of 5 $kBq$ each, located at the top
"right'' and at the bottom "left'' (from the view of the experimenter)
corners (see Fig. \ref{fig:2HS_original}). In Fig. \ref{fig:2HS_reconstructed},
the reconstructed activity distribution is shown. As before the same
number of data sets (10000) and CEM iteration steps ($k=10^{4}$ ) were used for the
calculation and the values shown are the arithmetic means for each
pixel. For the whole sample, the total estimated activity is $\lambda_{tot}=9954\:Bq$
with a standard deviation of $\sigma_{tot}=163\:Bq$. The lower and
upper limit of the 95\% symmetric coverage interval interval are $\lambda_{tot}^{\triangleleft}=9706\:Bq$
and $\lambda_{tot}^{\triangleright}=10241\:Bq$. The decision threshold
is $\lambda_{tot}^{\star}=42\:Bq$ and the detection limit $\lambda_{tot}^{\#}=90\:Bq$.
The integral efficiency over all detectors was calculated to be 31.6\%
for that activity distribution. The characteristic values of this
sample are summarized in Table \ref{tab:table_2HS}.

\begin{figure}
\begin{centering}
\includegraphics[scale=0.35]{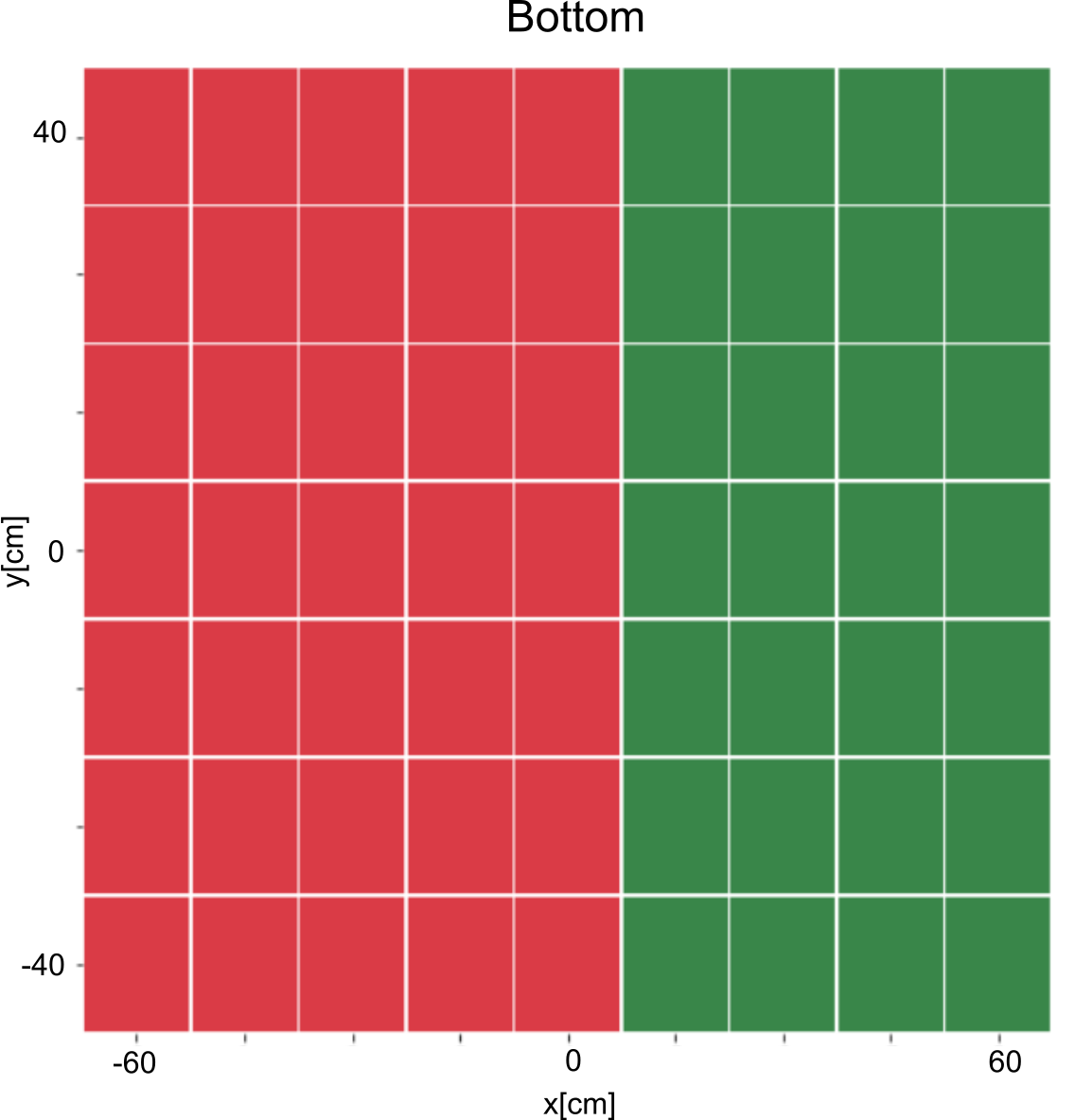}
\par\end{centering}
\caption{\label{fig:2HS_split}Splitting of sample n.2. Each of the three layers are split equally. The two blocks (red and green) are such that they have approximately the same number of cells}
\end{figure}

The sample was split into two blocks (as in Fig. \ref{fig:2HS_split}
where they are marked in red and green) which were analyzed separately.
The activity in Block 1 (red) is located in the bottom of the sample
whereas the activity in Block 2 (green) is located in the top. Due
to the fact that the lower source is better shielded (conveyor belt,
aluminum pallet, sample carrier box), the block containing this source
will clearly have a lower detection efficiency: indeed it has a total
efficiency of 27.1\% while the other block has an efficiency of 36.4\%.
Due to the difference in efficiency for block 1 and 2, the decision
thresholds were estimated to be respectively $\lambda_{1}^{\star}=71\:Bq$
and $\lambda_{2}^{\star}=64\:Bq$. These two values are larger than
the decision threshold for the whole sample due to the fact that,
when calculating the characteristic limits for a block, the radiation
generated by the other one needs to be considered as part of the background.
The detection limits are $\lambda_{1}^{\#}=163\:Bq$ and $\lambda_{2}^{\#}=150\:Bq$.
The mean activity in block 1 was calculated to be $\lambda_{1}=5124\:Bq$
with a standard deviation of $\sigma_{1}=284\:Bq$, whereas the for
block 2 a mean of $\lambda_{2}=4830\:Bq$ and standard deviation of
$\sigma_{2}=241\:Bq$ was estimated.The coverage intervals and all
other values are summarized in Table \ref{tab:table_2HS}.

\begin{table}
\begin{centering}
\begin{tabular}{|c|c|c|c|}
\cline{2-4} \cline{3-4} \cline{4-4} 
\multicolumn{1}{c|}{} & Total  & Block 1  & Block 2\tabularnewline
\hline 
Activity (Bq)  & 9954  & 5124  & 4830\tabularnewline
\hline 
Uncertainty (Bq)  & 163  & 284  & 241\tabularnewline
\hline 
Lower limit of coverage (Bq)  & 9706  & 4743  & 4383\tabularnewline
\hline 
Upper limit of coverage (Bq)  & 10241  & 5668  & 5159\tabularnewline
\hline 
Decision threshold (Bq)  & 42  & 71  & 64\tabularnewline
\hline 
Detection limit (Bq)  & 90  & 163  & 150\tabularnewline
\hline 
Efficiency (\%)  & 31.6  & 27.1  & 36.4\tabularnewline
\hline 
\end{tabular}
\par\end{centering}
\caption{\label{tab:table_2HS}Table with all the characteristic values of
the simulated sample n. 2 (activity distribution of Fig. \ref{fig:2HS_original})}
\end{table}

\begin{figure}
\begin{centering}
\includegraphics[scale=0.5]{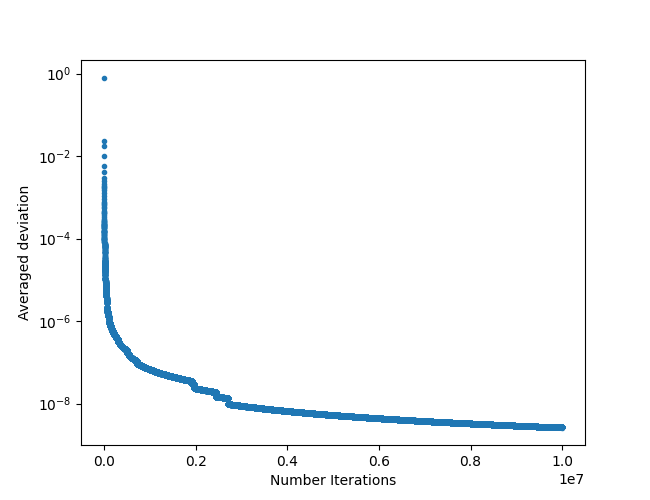}
\par\end{centering}
\begin{centering}
\includegraphics[scale=0.5]{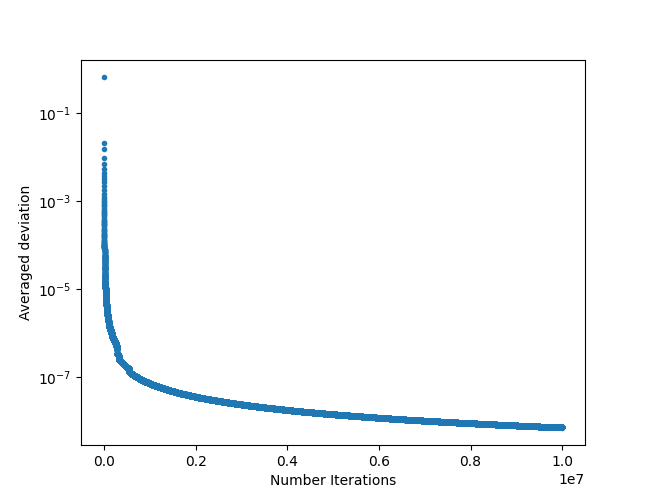} 
\par\end{centering}
\caption{\label{fig:plot_convergence}Plot in logarithmic scale of the average
deviation as a function of the iterative step $k$. The first picture on top shows the
plot for a single point source in the middle of the sample (as in Fig.
\ref{fig:1HS_original}); the second pictures shows the plot for two point sources
placed at opposite corners of the sample as in Fig. \ref{fig:2HS_original}}
\end{figure}

A convergence study of the CEM has also been performed. If we consider
the $\left(k+1\right)$-th step of the iterative CEM calculation of
formula (\ref{eq:recursive_formula_gross}), we can define the deviation
of cell $i$ from the previous iterative step 
\[
\Lambda_{i}^{k}=\left|\frac{\lambda_{i}^{k+1}-\lambda_{i}^{k}}{\lambda_{i}^{k}}\right|.
\]
If we average this quantity over all the cells, we have an indicator
\[
\Sigma^{k}=\frac{1}{N}\sum_{i=1}^{N}\Lambda_{i}^{k}
\]
of how much the algorithm has converged. In Fig. \ref{fig:plot_convergence}
we have plotted $\Sigma^{k}$ for the two cases of a point source
in the middle (as in Fig. \ref{fig:1HS_original}) and of two point
sources in opposite corners (as in Fig. \ref{fig:2HS_original}).

\section*{Conclusion}

In this paper we have presented an innovative approach to clearance measurements
of waste packages from nuclear facilities. The new procedure, borrowed
from medical physics, does not adopt algebraic evaluation models for
the estimate of the activity and goes one step further: the maximization
of the conditional entropy functional allows the estimate of the activity
distribution, cell by cell (as it happens for the image reconstruction).
The presented CEM method can be also adapted to the international
ISO 11929 standards by properly defining the characteristic limits
for the measured activities. Moreover, it allows to lower the characteristic
limits for the free release of the contaminated waste, e.g. by splitting
the sample in two or more parts: if a non-releasable sample containing
radioactive hotspots can be split into portions whose hotspots do
not exceed the legal thresholds separately, then we have achieved
the goal of reducing the amount of waste that needs to be stored in 
special repositories with consequent saving of money and resources. 

In this first study, we have then shown that the CEM method reconstructs
faithfully the radioactive distribution of a simulated metal block
with respectively one and two point sources. The calculation of the
characteristic limits of the whole block and of parts of the block
shows that we can significantly increase the amount of releasable
waste e.g. by splitting the sample. It was shown that the calculated
activity matches very well the original activity, easily within the
coverage interval. Furthermore, the localization and magnitude of
the activity was found to be in good agreement with the activity placed
in the sample both for a single hotspot and for two hotspots. The
described procedure to calculate characteristic limits provided reasonable
estimates of the decision threshold, of the detection limit as well as
of the coverage intervals for both the whole sample and for parts
of it. It was shown that the characteristic limits strongly depend
on the activity distribution, making clear the advantage of this approach.
Also we have found that the recursive algorithm requires a finite
number of iterations to converge within a given accuracy.

Further studies will include a quantitative analysis of the agreement
of the reconstructed image with the original sample, as well as a
general rule to fix the number of iterations needed to converge significantly.
Moreover we will report the results and conduct further analysis of
real samples. Another future extension of the method will include
techniques to improve the spatial resolution.

\section*{Acknowledgment}

We are thankful to prof. Rolf Michel for supporting our work and encouraging the writing of the present paper.


\appendix

\section{\label{sec:appendix_characteristic_limits}Definitions of the characteristic
values}

The formulae in section \ref{subsec:DIN_ISO} refer to the calculation of
characteristic values in case of a specific evaluation model and under
the assumption of a symmetric probability distribution of the measurands.
There are some cases though in which these assumptions don't hold
or it is even analytically impossible to determine the probability
distribution function and thus only a Monte Carlo sampling of the
measurand is possible \cite{characteristic_values}. If we define the probability distribution $f_{Y}\left(y|\mathbf{a}\right)$
that the measurand $Y$ takes the value $y$ given the set of conditions
$\mathbf{a}$ and a similar distribution$f_{Y}\left(y|\mathbf{a},y\geq0\right)$
of a positive measurand, we can build the cumulative distribution
function 
\[
F_{Y}\left(y|\mathbf{a}\right)=P\left(Y<y|\mathbf{a}\right)=\int_{-\infty}^{y}f_{Y}\left(\eta|\mathbf{a}\right)d\eta
\]
and also 
\[
I_{0}=\int_{0}^{\infty}f_{Y}\left(\eta|\mathbf{a};y\geq0\right)d\eta.
\]
With these definitions, we can generalize the expressions for all
the characteristic limits of the activity distribution given in section \ref{subsec:DIN_ISO}: 
\begin{itemize}
\item the mean or expectation value $\hat{y}$ is the best estimate for
the measurand 
\[
\hat{y}=E\left(Y|\mathbf{a},y\geq0\right)=I_{1}/I_{0};I_{1}=\int_{0}^{\infty}f_{Y}\left(y|\mathbf{a}\right)dy
\]
\item the standard uncertainty $u\left(\hat{y}\right)$ of the measurand
associated with the best estimate $\hat{y}$
\begin{multline*}
u\left(\hat{y}\right)=\sqrt{E\left(\left(Y-\hat{y}\right)^{2}|\mathbf{a},y\geq0\right)}=\sqrt{\frac{I_{2}}{I_{0}}-\hat{y}^{2}}\\
I_{2}=\int_{0}^{\infty}y^{2}f_{Y}\left(y|\mathbf{a}\right)dy
\end{multline*}
\item the decision threshold $y^{*}$ is the $\left(1-\alpha\right)$-quantile
of the $f_{Y}\left(y|\mathbf{\tilde{a}}\left(\tilde{y}=0\right)\right)$,
i.e. of the distribution of an assumed true value $\tilde{y}=0$.
It is calculated by solving 
\begin{equation}
1-F_{Y}\left(y^{*}|\mathbf{\tilde{a}}\left(\tilde{y}=0\right)\right)=\int_{y^{*}}^{\infty}f_{Y}\left(y|\mathbf{\tilde{a}}\left(\tilde{y}=0\right)\right)=\alpha.\label{eq:decision_threshold_appendix}
\end{equation}
This means that 
\begin{equation}
P\left(y>y^{*}|\tilde{y}=0\right)=\int_{y^{*}}^{\infty}f_{Y}\left(y|\mathbf{\tilde{a}}\left(\tilde{y}=0\right)\right)=\alpha\label{eq:decision_threshold_appendix_2}
\end{equation}
The decision threshold gives an indication whether the radioactivity
is present. If the outcome of a measurement exceeds $y^{*}$, the
measurand is potentially radioactive. If there is no real source,
the error committed by assuming that it is radioactive is called error
of the fist kind or false positive error. The probability of committing
such an error is exactly $\alpha$ (that is commonly chosen to be
0.95). If the measured value $y$ is below the decision threshold
$y^{*}$, the result cannot be attributed to the physical effect,
nevertheless it cannot be concluded that it is absent. If the physical
effect is really absent, the probability of taking the wrong decision,
that the effect is present, is equal to the specified probability,
$\alpha$ (probability of the wrong decision that it is not absent
if it actually is) 
\item the detection limit $y^{\#}$is the assumed true value of the measurand
if the decision threshold $y^{*}$ is the $\beta$-quantile of $f_{Y}\left(y|\mathbf{\tilde{a}}\left(\tilde{y}=y^{\#}\right)\right)$,
i.e. 
\begin{multline}
P\left(y<y^{*}|\tilde{y}=y^{\#}\right)=F_{Y}\left(y^{*}|\mathbf{\tilde{a}}\left(\tilde{y}=y^{\#}\right)\right)\\
=\int_{-\infty}^{y^{*}}f_{Y}\left(y|\mathbf{\tilde{a}}\left(\tilde{y}=y^{\#}\right)\right)=\beta.\label{eq:detection_limit_appendix}
\end{multline}
The detection limit is the lowest activity that can be measured with
a certain accuracy and with the applied measurement procedure. If
the threshold is not exceeded, the error committed if a source is
present is called error of the second kind or false negative error.
The probability of such an error is $\beta$ (that is commonly chosen
to be 0.95). The detection limit, $y^{\#}$, is the smallest true
value of the measurand, for which, the probability of the wrong decision,
that the physical effect is absent if it is not, does not exceed the
specified probability $\beta$. It is high enough compared to the
decision threshold $y^{*}$, that the probability of a false negative
decision does not exceed $\beta$ and is obtained as the smallest
solution of (\ref{eq:detection_limit_appendix}) 
\item the lower limit $y^{\vartriangleleft}$ and upper limit $y^{\vartriangleright}$
of the coverage interval are respectively the $\gamma/2$-quantile
and the $\left(1-\gamma/2\right)$-quantile of the density $f_{Y}\left(y|\mathbf{a},y\geq0\right)$
\begin{multline*}
F_{Y}\left(y^{\vartriangleleft}|\mathbf{a},y\geq0\right)=\frac{\gamma}{2}=\frac{I_{3}}{I_{0}}\\
I_{3}=\int_{0}^{y^{\vartriangleleft}}f_{Y}\left(y|\mathbf{a}\right)dy=F_{Y}\left(y^{\vartriangleleft}|\mathbf{a}\right)-F_{Y}\left(0|\mathbf{a}\right)
\end{multline*}
\begin{multline*}
1-F_{Y}\left(y^{\vartriangleright}|\mathbf{a},y\geq0\right)=\frac{\gamma}{2}=\frac{I_{4}}{I_{0}}\\
I_{4}=\int_{y^{\vartriangleright}}^{0}f_{Y}\left(y|\mathbf{a}\right)dy=1-F_{Y}\left(y^{\vartriangleright}|\mathbf{a}\right).
\end{multline*}
The coverage interval between the two limits defined above is such
that it contains the true value of the measurand with the coverage
probability $1-\gamma$ with $\gamma$ commonly chosen to be 0.05.
Using the Monte Carlo approach, the limits of the probabilistic
symmetric coverage interval $y^{\vartriangleleft}$ and $y^{\vartriangleright}$
are the $q_{\gamma/2}$ and the $q_{1-\gamma/2}$ quantiles of the
probability distribution $f_{Y}\left(\tilde{y}|\mathbf{a}\right)$
represented by the vector $\mathbf{y}_{M}=\left\{ y_{1},\ldots,y_{n_{M}}\right\} $
and taking into account that the measurand is non-negative. These
quantiles can be conveniently calculated from the vector $\mathbf{y}_{M}$
by searching the values $y_{k_{1}}=y^{\vartriangleleft}$ and $y_{k_{2}}=y^{\vartriangleright}$
with the conditions $y_{i}\geq0$, $k_{1}/n_{M,1}=\gamma/2$ amd $k_{2}/n_{M,1}=1-\gamma/2$
where $n_{M,1}$ is the number of elements in $\mathbf{y}_{M}$ that
are non-negative.
\end{itemize}

\end{document}